\documentstyle[preprint,revtex,eqsecnum]{aps}

\begin{document}

\begin{title}
Anatomy of the Soft-Photon Approximation in\\
Hadron-Hadron Bremsstrahlung
\end{title}

\author{M. K. Liou and Dahang Lin}

\begin{instit}
Department of Physics and Institute for Nuclear Theory\\
Brooklyn College of the City University of New York\\
Brooklyn, New York  11210
\end{instit}

\author{B. F. Gibson}

\begin{instit}
Theoretical Division\\
Los Alamos National Laboratory\\
Los Alamos, New Mexico  87545
\end{instit}

\begin{abstract}
A modified Low procedure for constructing soft-photon amplitudes has
been used
to derive two general soft-photon amplitudes, a two-s-two-t special
amplitude
$M^{TsTts}_{\mu}$ and a two-u-two-t special amplitude
$M^{TuTts}_{\mu}$, where
s, t and u are the Mandelstam variables. $M^{TsTts}_{\mu}$ depends
only on the
elastic T-matrix evaluated at four sets of (s,t) fixed by the
requirement that
the amplitude be free of derivatives ($\partial$T/$\partial$s and /or
$\partial$T/$\partial t$). Likewise $M^{TuTts}_{\mu}$ depends only on
the
elastic T-matrix evaluated at four sets of (u,t) also fixed by the
requirement
that the amplitude $M^{TuTts}_{\mu}$ be free of derivatives
($\partial$T/$\partial$u and/or $\partial$T/$\partial t$). In
deriving these
two amplitudes, we imposed the condition that $M^{TsTts}_{\mu}$ and
$M^{TuTts}_{\mu}$ reduce to $\bar{M}^{TsTts}_{\mu}$ and
$\bar{M}^{TuTts}_{\mu}$, respectively, their tree level
approximations. The
amplitude $\bar{M}^{TsTts}_{\mu}$ represents photon emission from a
sum of
one-particle t-channel exchange diagrams and one-particle s-channel
exchange
diagrams, while the amplitude $\bar{M}^{TuTts}_{\mu}$ represents
photon
emission from a sum of one-particle t-channel exchange diagrams and
one-particle u-channel exchange diagrams. The precise expressions for
$\bar{M}^{TsTts}_{\mu}$ and $\bar{M}^{TuTts}_{\mu}$ are determined by
using the
radiation decomposition identities of Brodsky and Brown. We also
demonstrate
that two Low amplitudes $M^{Low(st)}_{\mu}$ and $M^{Low(ut)}_{\mu}$,
derived
using Low's standard procedure, can be obtained from
$M^{TsTts}_{\mu}$ and
$M^{TuTts}_{\mu}$, respectively, as an expansion in powers of K
(photon energy)
when terms of order K and higher are neglected. We point out that it
is
theoretically impossible to describe all nuclear bremsstrahlung
processes by
using only a single class of soft-photon amplitudes. At least two
different
classes are required: the amplitudes (such as $M^{TsTts}_{\mu}$,
$M^{Low(st)}_{\mu}$ and $\bar{M}^{TsTts}_{\mu}$) which depend on s
and t or the
amplitudes (such as $M^{TuTts}_{\mu}$, $M^{Low(ut)}_{\mu}$ and
$\bar{M}^{TuTts}_{\mu}$) which depend on u and t. When resonance
effects are
important, the amplitude $M^{TsTts}_{\mu}$, not $M^{Low(st)}_{\mu}$,
should be
used. For processes with strong u-channel exchange effects, the
amplitude
$M^{TuTts}_{\mu}$ should be the first choice. As for those processes
which
exhibit neither s-channel resonance effects nor u-channel exchange
effects, all
amplitudes converge essentially to the same description. Finally, we
discuss
the relationship between the two classes.
\end{abstract}

\pacs{ }

\section{INTRODUCTION}\label{sec:1}
Hadron-hadron bremsstrahlung processes have attracted much attention
during the
last three decades. Processes like nucleon-nucleon bremsstrahlung
(pp$\gamma$
and np$\gamma$) [1-3], proton-deuteron bremsstrahlung (pd$\gamma$)
[4,5],
proton-helium bremsstrahlung (p$\alpha\gamma$) [4,6], proton-carbon
bremsstrahlung (p$^{12}$C$\gamma$) [7], proton-oxygen bremsstrahlung
(p$^{16}$O$\gamma$) [8], and pion-proton bremsstrahlung
($\pi^{\pm}$p$\gamma$)
[9] are the best-known examples, because they have been studied both
experimentally and theoretically. There exist a variety of reasons
for
investigating these processes: (i) One of the important goals is the
investigation of off-shell effects in the scattering amplitude. For
instance,
the pp$\gamma$ and np$\gamma$ processes have been extensively studied
since
1963 to investigate the off-shell behavior of two-nucleon
interactions. Most
theoretical studies have focused on nonrelativistic potential model
calculations using various phenomenological potentials as input, with
the goal
that the best potential could be selected from comparison with
pp$\gamma$
and/or np$\gamma$ data [1-3]. Recently, the observation of energetic
photons
from heavy-ion collisions has created a growing interest in
understanding the
basic production mechanism of these high energy photons [10]. The
np$\gamma$
process has received renewed attention because it appears to be the
most likely
source of such energetic photons. Moreover, np$\gamma$ is probably an
ideal
process for studying meson exchange effects [11,12]. (ii)
Bremsstrahlung
processes have been used as a tool to investigate electromagnetic
properties of
resonances. The most successful example is the determination of the
magnetic
moment of the $\Delta^{++}$($\Delta^{0}$) from the $\pi^{+}$p$\gamma$
($\pi^{-
}$p$\gamma$) data in the energy region of the $\Delta$(1232)
resonance [9,13].
(iii) The study of nucleon-nucleus and nucleus-nucleus bremsstrahlung
processes
in the vicinity of resonances, such as the p$^{12}$C$\gamma$ process
near the
1.7 and 0.5 MeV resonances [7] or the p$^{16}$O$\gamma$ process near
the 2.66
MeV resonance [8], was originally suggested for investigating details
of
nuclear reactions. Such bremsstrahlung measurements can be used to
extract the
nuclear time delay, and the time delay can be used to distinguish
between a
direct nuclear reaction and a compound nuclear reaction. That
bremsstrahlung
emission can be used as a tool to measure time delay has been
confirmed
experimentally: three separate experimental groups have measured the
p$^{12}$C$\gamma$ cross sections and then used these cross sections
to extract
nuclear time delays [7]. (iv) Testing theoretical models and
approximations has
been another important aspect of studying hadron-hadron
bremsstrahlung
processes, especially those processes containing significant
resonance or
exchange effects. The combined experimental and theoretical
investigations of
the $\pi^{\pm}$p$\gamma$ and p$^{12}$C$\gamma$ processes led to a
surprising
conclusion [14,15]: these cross sections cannot be described by the
conventional soft-photon amplitudes (evaluated at a single energy and
scattering angle) which had been the standard since 1958 when Low
first derived
them. They fail completely to fit the experimental data. These
observations
indicate why the study of bremsstrahlung processes with significant
resonance
effects or meson exchange effects can provide a sensitive test of
theoretical
models and approximations.

Among the various models and approximations proposed during the past
three
decades for bremsstrahlung calculations, the best-known approximation
is the
soft-photon approximation. This approximation is based upon a
fundamental
theorem: the soft-photon theorem or the low-energy theorem for
photons. The
theorem was first derived by Low [16]; it was generalized and
extended later by
many other authors [17,14,9]. Various soft-photon amplitudes, which
are
consistent with the theorem, have been constructed by using the
standard Low
procedure [16]. This involves the following steps: (a) Obtain the
external
amplitude, $M^{(E)}_{\mu}$, from the four external emission diagrams
and expand
$M^{(E)}_{\mu}$ in powers of the photon energy K. (b) Impose the
gauge
invariant condition, $M^{(I)}_{\mu}$K$^{\mu}$ =
$-M^{(E)}_{\mu}$K$^{\mu}$, to
obtain the leading term (order K$^{0}$) of the internal emission
amplitude,
$M^{(I)}_{\mu}$. (c) Combine $M^{(E)}_{\mu}$ and $M^{(I)}_{\mu}$ to
obtain the
total bremsstrahlung amplitude, $M^{(T)}_{\mu}$. Low's soft-photon
amplitude
$M^{Low}_{\mu}$, which is independent of off-shell effects, is
defined by the
first two terms of the expansion of $M^{(T)}_{\mu}$. A universal
feature of all
soft-photon amplitudes is that they depend only on the corresponding
elastic
amplitude and electromagnetic constants of the participating
particles.
Therefore, the soft-photon approximation is referred to as the
on-shell
approximation, and the calculations based on the soft-photon
approximation are
classified as model-independent.

The reader will note that the standard procedure cannot be used to
obtain an
internal contribution which is separately gauge invariant [9,18].
Therefore,
it is difficult to obtain a general form for the internal amplitude
by using
the standard procedure. In order to derive the general soft-photon
amplitude, a
modified Low procedure was proposed recently [9,18]. The modified
procedure
includes four steps. Because the determination of the general
amplitude
$M_{\mu}$ is guided by the derivation for the corresponding special
amplitude
$\bar{M}_{\mu}$ which can be rigorously derived at the tree level, we
first
apply the modified procedure to find $\bar{M}_{\mu}$: (1) Obtain the
external
amplitude $\bar{M}^{E}_{\mu}$ from a set of tree level external
diagrams. (2)
Obtain the internal contribution $\bar{M}^{I}_{\mu}$, which
represents photon
emission from a dominant internal line (or lines), and split
$\bar{M}^{I}_{\mu}$ into four 1 amplitudes by using the radiation
decomposition
identities of Brodsky and Brown [19]. (3) Obtain an additional gauge
invariant
term $\bar{M}^{G}_{\mu}$ by imposing the gauge invariant condition,
$\bar{M}^{G}_{\mu}K^{\mu}$ = $-\bar{M}^{EI}_{\mu}K^{\mu}$. Here
$\bar{M}^{EI}_{\mu}$ is the sum of $\bar{M}^{E}_{\mu}$ and
$\bar{M}^{I}_{\mu}$:
$\bar{M}^{EI}_{\mu}$ = $\bar{M}^{E}_{\mu}$ + $\bar{M}^{I}_{\mu}$. (4)
Combine
$\bar{M}^{EI}_{\mu}$ and $\bar{M}^{G}_{\mu}$ to obtain the total
amplitude,
$\bar{M}_{\mu}$ = $\bar{M}^{EI}_{\mu}$ + $\bar{M}^{G}_{\mu}$. The
amplitude
$\bar{M}_{\mu}$, especially the expression for $\bar{M}^{I}_{\mu}$,
can then be
used to determine the general amplitude $M_{\mu}$ by applying the
modified
procedure again: (1) Obtain the external amplitude $M^{E}_{\mu}$ from
four
general external diagrams. (This step is identical to the first step
of the
standard procedure.) (2) Construct an internal contribution
$M^{I}_{\mu}$ which
reduces to $\bar{M}^{I}_{\mu}$ at the tree level approximation. (3)
Obtain an
additional gauge invariant term $M^{G}_{\mu}$ by imposing the gauge
invariant
condition, $M^{G}_{\mu}K^{\mu} = -M^{EI}_{\mu}K^{\mu}$. Here,
$M^{EI}_{\mu}=
M^{E}_{\mu} + M^{I}_{\mu}$. (4) Combine $M^{EI}_{\mu}$ with
$M^{G}_{\mu}$ to
obtain the total amplitude, $M_{\mu} = M^{EI}_{\mu} + M^{G}_{\mu}$,
which
should reduce to $\bar{M}_{\mu}$ at the tree level approximation. The
first two
terms of the expansion of $M_{\mu}$, which can be written in terms of
the
complete elastic T-matrix and electromagnetic constants of the
participating
particles, define a general soft-photon amplitude. Here, the
expansion of
$M_{\mu}$ is performed in such a way that the expanded $M_{\mu}$ will
depend on
the elastic T-matrix, evaluated for two Mandelstam variables, but it
will be
free of any derivative of the T-matrix with respect to those two
specified
Mandelstam variables.

The purpose of this work is to study the soft-photon approximation
systematically. We apply both the standard procedure and the modified
procedure
to derive various soft-photon amplitudes, which fall naturally into
two classes
delineated by the choice of Mandelstam variables. We find that one of
these two
classes is completely new; it has been totally ignored in the
literature. We
show that these two classes are independent and they are equally
important for
describing bremsstrahlung processes. In order to make our point more
precisely,
let us consider photon emission accompanying the scattering of two
particles A
and B (s-channel reaction):
\begin{equation}
A(q_{i}^{\mu}) + B(p_{i}^{\mu}) \longrightarrow A(q_{f}^{\mu}) +
B(p_{f}^{\mu}) + \gamma(K^{\mu}).
                                                        \label{eq:1}
\end{equation}
\noindent Here, $q_{i}^{\mu}(q_{f}^{\mu})$ and $p_{i}^{\mu}(p_{f}^{\mu})$
are the initial
(final) four-momenta for particles A and B, respectively, and
$K^{\mu}$ is the
four-momentum for the emitted photon with polarization
$\epsilon^{\mu}$. We
assume that the particle A has mass m$_{A}$ and charge Q$_{A}$ while
the
particle B has mass m$_{B}$ and charge Q$_{B}$. For simplicity, we
also assume
that both A and B are spinless particles since our problem does not
depend on
the spin of the participating particles. From the process (1), we can
define
the following Mandelstam variables: $s_{i}$ = ($q_{i}$ +
$p_{i}$)$^{2}$,
$s_{f}$ = ($q_{f}$ + $p_{f}$)$^{2}$, $t_{p}$ = ($p_{f} -
p_{f}$)$^{2}$, $t_{q}$
= ($q_{f} - q_{f}$)$^{2}$, $u_{1}$ = ($p_{f} - q_{i}$)$^{2}$, and
$u_{2}$ =
($q_{f} - p_{i}$)$^{2}$. Since a soft-photon amplitude depends only
upon two
independent variables, chosen from the above possible Mandelstam
variables, we
can express the two independent classes of soft-photon amplitude as
$M^{(1)}_{\mu}$(s, t) and $M^{(2)}_{\mu}$(u, t). Here,
$M^{(1)}_{\mu}$(s, t)
includes all amplitudes which depend upon s [choosing from
s$_{i}$,
s$_{f}$ and other linear combinations s$_{\alpha\beta}$=($\alpha
s_{i}$ +
$\beta s_{f}$)/($\alpha + \beta$), $\alpha\neq$ 0 and $\beta\neq$ 0]
and t
[choosing from t$_{p}$, t$_{q}$ and other linear combinations
t$_{\alpha '\beta '}$=($\alpha '$t$_{p}$ +
$\beta '$t$_{q}$)
/($\alpha ' + \beta '$), $\alpha ' \neq$ 0 and
$\beta ' \neq$ 0]
while $M^{(2)}_{\mu}$(u, t) involves all those amplitudes which
depend upon u
[choosing from u$_{1}$, u$_{2}$ and other combinations
u$_{\bar{\alpha}\bar{
\beta}}$ = ($\bar{\alpha}u_{1} + \bar{\beta}u_{2}$)/$\bar{\alpha} +
\bar{\beta}$), $\bar{\alpha}\neq$0 and $\bar{\beta}\neq$0] and t.

The first class, $M^{(1)}_{\mu}$(s, t), contains three interesting
amplitudes:
(i) the conventional Low amplitude $M^{Low(st)}_{\mu}$ ($\bar{s},
\bar{t}$), (ii) the Feshbach-Yennie amplitude $M^{FY}_{\mu}$(s$_{i}$,
s$_{f}$;
t), and (iii) the two-s-two-t special amplitude
$M^{TsTts}_{\mu}(s_{i}$,
$s_{f}$; $t_{p}, t_{q}$) [or the special two-energy-two-angle
amplitude
$M^{TETAS}_{\mu}(s_{i}$, $s_{f}$; $t_{p}, t_{q}$)]. The
$M^{Low(st)}_{\mu}$
amplitude can be derived using the standard Low procedure. Since this
latter
amplitude depends on $\bar{s} = s_{11} = (s_{i} + s_{f})$/2 and
$\bar{t} =
t_{11} = (t_{p} + t_{q})$/2, it is a one-energy-one-angle (OEOA)
amplitude
[14]. $M^{Low(st)}_{\mu}$ has been widely used to calculate cross
sections for
many bremsstrahlung processes for more than thirty years. However,
recent
investigations have shown that it fails to describe those
bremsstrahlung
processes which are dominated by resonance effects. The
Feshbach-Yennie
amplitude is a special two-energy-one-angle amplitude [14]. It is
interesting
primarily because it was the first soft-photon amplitude which was
used to
describe some bremsstrahlung processes with scattering resonances and
to
extract the nuclear time delay from bremsstrahlung cross sections.
The
amplitude $M^{TsTts}_{\mu}$, as we will see later in this work, is
the general
amplitude in the $M^{(1)}_{\mu}$(s, t) class since all other
amplitudes, such
as $M^{Low(st)}_{\mu}$ and $M^{FY}_{\mu}$, can be reproduced from it.
Because
the modified procedure is used to derive $M^{TsTts}_{\mu}$, the
amplitude will be shown to be
independent of the derivatives of the elastic T-matrix with respect
to s or t.
The amplitude has been tested experimentally. The amplitude
$M^{TETAS}_{\mu}$,
which is a practical version of $M^{TsTts}_{\mu}$, can be used to
describe
almost all available $\pi^{\pm}p\gamma$ and $p^{12} C\gamma$ data. It
has been
used to determine the magnetic moments of $\Delta^{++}$ and
$\Delta^0$ from the
$\pi^+ p\gamma$ and $\pi^- p\gamma$ data, respectively. Although the
$M^{TETAS}_{\mu}$ amplitude should be used, it has never actually
applied to
extract the nuclear time delay from the bremsstrahlung data.

The second class, $M^{(2)}_{\mu}$(u, t), is completely new. It has
not been
previously studied or discussed in the literature. Here, we show (i)
how the
standard procedure can be used to derive another Low's amplitude,
$M^{Low(ut)}_{\mu}$ (u$_{11}$,t$_{11}$) where u$_{11}$=(u$_1$ +
u$_2)$/2, and
(ii) how the modified procedure can be used to obtain the
general
amplitude for the second class, the two-u-two-t special amplitude
$M^{TuTts}_{\mu}($u$_1$, u$_{2}$; t$_p$, t$_q$). We explain why we
expect
$M^{TuTts}_{\mu}$ to play a major role in predicting and describing
those
processes which are dominated by exchange current effects.

We also discuss the relationship between $M^{(1)}_{\mu}$(s, t) and
$M^{(2)}_{\mu}$(u, t). In particular, we show that the two classes
can be
interchanged, $M^{(1)}_{\mu}$ (s, t) $\longleftrightarrow$
$M^{(2)}_{\mu}$ (u,
t), if Q$_B$ is replaced by $-$Q$_B$, Q$_B$ $\longrightarrow$
$-$Q$_B$, and
p$^\mu_i$ is interchanged with $-$p$^\mu_f$, p$^\mu_i$
$\longleftrightarrow$
$-$p$^\mu_f$.

\section{ELASTIC SCATTERING T-MATRIX}\label{sec:2}

The bremsstrahlung process which we wish to study is given by Eq.\
(\ref{eq:1}). The five four momenta in Eq.\ (\ref{eq:1})
satisfy energy-momentum conservation:

\begin{equation}
q_{i}^{\mu} + p_{i}^{\mu} = q_{f}^{\mu} + p_{f}^{\mu} + K^{\mu}.
                                                         \label{eq:2}
\end{equation}
\noindent In the limit when K approached zero, the bremsstrahlung
process reduces to the
corresponding A-B elastic scattering process,

\begin{equation}
A(q_{i}^{\mu}) + B(p_{i}^{\mu}) \longrightarrow A(\bar{q}_{f}^{\mu})
+ B(\bar{p}_{f}^{\mu}),
                                                        \label{eq:3}
\end{equation}
\noindent where
\begin{mathletters}
\begin{equation}
\bar{q}_{f}^{\mu} = \lim_{K \rightarrow 0} q_{f}^{\mu}
                                                        \label{eq:4a}
\end{equation}
\noindent and
\begin{equation}
\bar{p}_{f}^{\mu} = \lim_{K\rightarrow 0} p_{f}^{\mu}
                                                        \label{eq:4b}
\end{equation}
\end{mathletters}

\noindent The energy-momentum conservation relation defined in
Eq.\ (\ref{eq:2}) becomes
\begin{equation}
q_{i}^{\mu} + p_{i}^{\mu} = \bar{q}_{f}^{\mu} + \bar{p}_{f}^{\mu}
                                                        \label{eq:5}
\end{equation}

A diagram which represents the A-B elastic scattering process is
shown in Fig.~1(a).
In this diagram, $\bar{T}$ represents the A-B elastic
scattering
T-matrix. $\bar{T}$ is an on-shell T-matrix because all four external
lines
(legs) are on their mass shells. For the bremsstrahlung process,
which will be
discussed in the next section, the exact bremsstrahlung amplitude
(without the
soft-photon approximation) involves half-off-shell T-matrices. Each
of these
T-matrices, on-shell or half-off-shell, can be written in terms of
six Lorentz
invariants, chosen from $s(s_{i}$ or $s_{f}),\ t(t_{p}$ or $t_{q}),\,
u(u_{1}$ or $u_{2}),\, q'$$^{2}_{f} [q^{2}_{f}$ or $\Delta_{a} = (q_{f} +
K)^{2}],
q'$$^{2}_{i}[q^{2}_{i}$ or $\Delta_{b} = (q_{i} - K)^{2}],\
p'$$^{2}_{f}[p^{2}_{f}$ or $\Delta_{c} = (p_{f} + K)^{2}0]$, and
$p'$$^{2}_{i}[p^{2}_{i}$ or $\Delta_{d} = (p_{i} - K)^{2}]$.
\noindent Thus,
any T-matrix can be written as
\begin{mathletters}
\begin{equation}
T(s, t, q'$$^{2}_{i}, p'$$^{2}_{i}, q'$$^{2}_{f}, p'$$^{2}_{f})
                                                       \label{eq:6a}
\end{equation}
\noindent or
\begin{equation}
T(u, t, q'$$^{2}_{i}, p'$$^{2}_{i}, q'$$^{2}_{f},
p'$$^{2}_{f}).
                                                       \label{eq:6b}
\end{equation}
\end{mathletters}
\noindent As in the examples, let us define the following T-matrices
which will be used later.

(i) The elastic (on-shell) T-matrix can be written as a function of
two
independent variables; e.g.,
\begin{mathletters}
\begin{equation}
T(s, t) \equiv T(s, t, m^2_A, m^2_B, m^2_A, m^2_B)      \label{eq:7a}
\end{equation}
\noindent or
\begin{equation}
T(u, t) \equiv T(u, t, m^2_A, m^2_B, m^2_A, m^2_B).     \label{eq:7b}
\end{equation}
\end{mathletters}
\noindent This is because $q^2_i, p^2_i, q^2_f$ and $p^2_f$ satisfy
the on-mass-shell conditions,
\begin{mathletters}
\begin{eqnarray}
q^2_i &=& m^2_A, \nonumber \\
p^2_i &=& m^2_B, \nonumber \\
q^2_f &=& \bar{q}^{2}_{f} = m^2_A,
                                                      \label{eq:8a}
\end{eqnarray}
and
\[
p^2_f = \bar{p}^2_f = m^2_B,
\]
\noindent and only two of the three Mandelstam variables are
independent since they satisfy the following condition:
\begin{equation}
s + t + u = 2m^{2}_{A} + 2m^{2}_{B},
                                                      \label{eq:8b}
\end{equation}
\end{mathletters}
\noindent where
\begin{eqnarray}
s &=& (q_{i} + p _{i})^{2} = (\bar{q}+\bar{p}_{f})^{2},
                                                      \nonumber \\
t &=& (\bar{p}_{f}-p_{i})^{2} = (\bar{q}_{f}-q_{i})^{2},
                                                      \label{eq:9}
\end{eqnarray}
and
\[
u = (\bar{q}_{f}-p_{i})^{2}\ = (\bar{p}_{f}-q_{i})^{2}.
\]

(ii) Five diagrams which represent the bremsstrahlung process (1.1)
are shown
in Fig.~2. A half-off-shell T-matrix can be defined if a photon of
momentum
$K^\mu$ is emitted from the outgoing A-particle [see Fig.~2(a)]. This
T-matrix
can be written as a function of three independent variables,
\begin{mathletters}
\begin{equation}
T(s_{i}, t_{p}, \Delta_{a}) \equiv T(s_{i}, t_{p}, m^{2}_{A},
m^{2}_{B}, \Delta_{a}, m^{2}_{B})                      \label{eq:10a}
\end{equation}
\noindent or
\begin{equation}
T(u_{1}, t_{p}, \Delta_{a}) \equiv T(u_{1}, t_{p}, m^{2}_{A},
m^{2}_{B}, \Delta_{a}, m^{2}_{B}),
                                                       \label{eq:10b}
\end{equation}
\end{mathletters}
\noindent where
\begin{eqnarray}
s_{i} &=& (q_{i}+ p_{i})^{2},\nonumber \\
t_{p} &=& (p_{f}- p_{i})^{2},
                                                     \label{eq:11}\\
u_{1} &=& (p_{f}- q_{i})^{2},\nonumber
\end{eqnarray}
\noindent and
\[
\Delta_{a} \equiv (q_{f}+ K)^{2} = m^{2}_{A} + 2 q_{f}\cdot K.
\]
\noindent It is easy to show that

\begin{equation}
s_{i} + t_{p} + u_{1} = \Delta_{a} + m^{2}_{A} + 2m^{2}_{B}.
                                                       \label{eq:12}
\end{equation}

(iii) A half-off-shell T-matrix can be defined if a photon of
momentum $K^\mu$
is emitted from the incoming A-particle [see Fig.~2(b)]. This T-matrix
is a
function of three independent variables.
\begin{mathletters}
\begin{equation}
T(s_{f}, t_{p}, \Delta_{b}) \equiv T( s_{f}, t_{p}, \Delta_{b},
m^{2}_{B}, m^{2}_{A}, m^{2}_{B})                      \label{eq:13a}
\end{equation}
or
\begin{equation}
T(u_{2}, t_{p}, \Delta_{b}) \equiv T( u_{2}, t_{p}, \Delta_{b},
m^{2}_{B}, m^{2}_{A}, m^{2}_{B})
                                                      \label{eq:13b}
\end{equation}
\end{mathletters}
where
\begin{eqnarray}
s_{f}&=& (q_{f}+ p_{f})^{2},                          \nonumber \\
u_{2}&=& (q_{f}- p_{i})^{2},
                                                     \label{eq:14}
\end{eqnarray}
and
\[
\Delta_{b} \equiv (q_{i}- K)^{2} = m^{2}_{A} - 2 q_{i}\cdot K.
                                                          \nonumber
\]
We can show that
\begin{equation}
s_{f} + t_{p} + u_{2} = \Delta_{b} + m^{2}_{A} + 2m^{2}_{B}.
                                                     \label{eq:15}
\end{equation}

(iv) A half-off-shell T-matrix can be defined if a photon of momentum
$K^\mu$
is emitted from the outgoing B-particle [see Fig.~2(c)]. This T-matrix
is a
function of three independent variables,
\begin{mathletters}
\begin{equation}
T(s_{i}, t_{q}, \Delta_{c}) \equiv T(s_{i}, t_{q}, m^{2}_{A},
m^{2}_{B}, m^{2}_{A}, \Delta_{c})                   \label{eq:16a}
\end{equation}
or
\begin{equation}
T(u_{2}, t_{q}, \Delta_{c}) \equiv T(u_{2}, t_{q}, m^{2}_{A},
m^{2}_{B}, m^{2}_{A}, \Delta_{c})
                                                    \label{eq:16b}
\end{equation}
\end{mathletters}
where
\begin{equation}
t_{q}= (q_{f}- q_{i})^{2},
                                                    \label{eq:17}
\end{equation}
and
\[
\Delta_{c} \equiv (p_{f}+ K)^{2} = m^{2}_{B} + 2 p_{f}\cdot
K.\nonumber
\]
The following relation can be easily proved:
\begin{equation}
s_{i} + t_{q} + u_{2} = \Delta_{c} + m^{2}_{A} + 2m^{2}_{A}.
                                                    \label{eq:18}
\end{equation}
(v) A half-off-shell T-matrix can be defined if a photon of momentum
$K^\mu$ is
emitted from the incoming B-particle [see Fig.~2(d)]. This T-matrix is
a function
of three independent variables,
\begin{mathletters}
\begin{equation}
T(s_{f}, t_{q}, \Delta_{d}) \equiv T(s_{f}, t_{q}, m^{2}_{A},
\Delta_{d}, m^{2}_{A}, m^{2}_{B})                       \label{eq:19a}
\end{equation}
\noindent or
\begin{equation}
T(u_{1}, t_{q}, \Delta_{d}) \equiv T(u_{1}, t_{q}, m^{2}_{A},
\Delta_{d}, m^{2}_{A}, m^{2}_{B})
                                                        \label{eq:19b}
\end{equation}
\end{mathletters}
\noindent where
\begin{equation}[
\Delta_{d}\equiv (p_{i}- K)^{2} = m^{2}_{B} - 2 p_{i}\cdot K.
                                                        \label{eq:20}
\end{equation}
\noindent It is not difficult to prove that
\begin{equation}
s_{f} + t_{q} + u_{1} = \Delta_{d} + m^{2}_{B} + 2m^{2}_{A}.
                                                       \label{eq:21}
\end{equation}

The above discussion illustrates clearly that there are at least two
different
ways of choosing independent variables for each T-matrix. The first
choice
involves s and t while the second choice involves u and t. In the
case that one
is dealing with the exact amplitude for bremsstrahlung (in contrast
to the
soft-photon approximation which is the subject of this paper), these
two
choices must be equivalent. {\it However, we shall see below that if
one
soft-photon amplitude is parametrized in terms of s and t and another
soft-photon amplitude is parametrized in terms of u and t, then the
two
amplitudes are no longer equivalent \rm}. The soft-photon
approximation makes
the two resulting amplitudes different. Which independent variables
to select
and how to parametrize T-matrices in terms of them is an important
problem
which must be carefully considered in order to establish the optimal
soft-photon amplitude for specific bremsstrahlung processes. Since
the elastic
scattering diagrams serve as the ultimate source graphs from which
all
bremsstrahlung diagrams are generated, the independent variables in a
soft-photon amplitude are specified by the choice of independent
variables made
in expressing the elastic T-matrix.

In order to illustrate this point, let us consider two special elastic
scattering cases. In each case, we assume that the elastic scattering
process is determined by a set of one-particle exchange diagrams.
The first case is
depicted in Fig.~1(b) and the second case in Fig.~1(c).

In the case shown in Fig.~1(b), the elastic A-B scattering
process is
determined by a sum of one-particle t-channel exchange diagrams and
one-particle s-channel exchange diagrams. In other words, we assume
that the
A-B system involves the t-channel exchange of particles and the
s-channel
exchange of particles (an intermediate state or scattering
resonance). The
one-particle s-channel exchange diagrams are the dominant elastic
diagrams in
the resonance regions. [Two well-known examples are $\pi$N scattering
in the
$\Delta$(1232) resonance region and p$^{12}$C scattering near either
the 1.7
MeV resonance or the 0.5 MeV resonance.]  The elastic scattering
T-matrix
corresponding to Fig.~1(b) has the form:

\begin{equation}
\bar{T}(s, t) = \bar{T}_{C}(t) + \bar{T}_{D}(s),
                                                         \label{eq:22}
\end{equation}
where
\begin{mathletters}
\begin{equation}
\bar{T}_{C}(t) = \sum_{n} \Gamma^{AC}_{n} \frac{i}{t-(m^{C}_{n})^{2}
+ i\epsilon}\Gamma^{CB}_{n}
                                                        \label{eq:23a}
\end{equation}
and
\begin{equation}
\bar{T}_{D}(s) = \sum_{l}\Gamma^{ADB}_{l} \frac{i}{s-(m^{D}_{l})^{2}
+ i\epsilon}\Gamma^{ADB}_{l}.
                                                       \label{eq:23b}
\end{equation}
\end{mathletters}
In Eqs.\ (\ref{eq:23a}) and (\ref{eq:23b}), m$^C_n$ (n = 1, 2, $\ldots$)
are the masses of the t-channel exchange particles C$_n$, $\Gamma^{AC}_n$
are the A-C$_n$-A vertices,
$\Gamma^{CB}_n$ are the B-C$_n$-B vertices, m$^D_l$ ($l$ = 1, 2,
$\ldots$) are the masses of the intermediate particles (s-channel
exchange particles) D$_l$, $\Gamma^{ADB}_l$ are the A-D$_l$-B
vertices, and s and t are defined by Eq.\ (\ref{eq:9}).
Conservation of charge requires that all
t-channel exchange particles C$_n$ be neutral and the charge of every
s-channel
exchange particle D$_l$ must be Q$_A$ + Q$_B$. If these diagrams are
used as
source graphs to describe internal emission, t-channel exchange
particles make
no contribution to internal emission because they have no charge.
Therefore,
photon emission from the s-channel exchange determines the entire
internal
amplitude in this case.

In the second case, as shown in Fig.~1(c), the elastic A-B
scattering
process is determined by a sum of one-particle u-channel exchange
diagrams. In
other words, we assume that the A-B system involves the t-channel
exchange
particles and the u-channel exchange particles F$_j$ (j = 1, 2,
$\ldots$). The elastic scattering T-matrix corresponding to
Fig.~1(c) has the form:

\begin {equation}
\bar{T}(t, u) = \bar{T}_{C}(t) + \bar{T}_{F}(u),       \label{eq:24}
\end{equation}
\noindent where $\bar{T}_{C}$(t) is given by Eq.\ (\ref{eq:23a}) and
\begin{equation}
\bar{T}_{F}(u)= \sum_{j} \Gamma^{AFB}_{j} \frac{i}{u-(m^{F}_{j})^{2}
+ i\epsilon}\Gamma^{AFB}_{j}.
                                                      \label{eq:25}
\end{equation}
\noindent In Eq.\ (\ref{eq:25}), m$^F_j$ (j = 1, 2, $\ldots$) are the
masses of
the u-channel exchange particles F$_j$, $\Gamma^{AFB}_j$ are the
A-F$_j$-B
vertices, and u is defined by Eq.\ (\ref{eq:9}). The charge of every
u-channel
exchange particle is $Q_{A} - Q_{B}$. If $Q_{A} - Q_{B} \neq 0$, then
photon
emission from the u-channel exchange particles determines the entire
internal
amplitude in this case.\\

\section{BREMSSTRAHLUNG AMPLITUDE AT THE TREE LEVEL}\label{sec:3}
\subsection{Photon emission from the tree diagrams given by Fig.~1(b)}
\indent If the elastic scattering diagrams given by Fig.~1(b) are
used as
source graphs to generate bremsstrahlung diagrams, then we obtain
Fig.~3. Figs.~3(a)-3(d) represent the external emission diagrams
and Fig.~3(e) represents the internal emission diagram.
The external bremsstrahlung amplitude corresponding
to Figs.~3(a)-3(d) has the form [21]

\begin{eqnarray}
\bar{M}^{E(CD)}_{\mu} &=& Q_{A}\frac{q_{f\mu}}{q_{f}\cdot
K}\bar{T}(s_{i}, t_{p}) - Q_{A}\frac{q_{i\mu}}{q_{i}\cdot
 K}\bar{T}(s_{f},t_{p})                            \nonumber \\
&+& Q_{B}\frac{p_{f\mu}}{p_{f}\cdot K}\bar{T}(s_{i}, t_{q}) -
Q_{B}\frac{p_{i\mu}}{p_{i}\cdot K}\bar{T}(s_{f},t_{q}),
                                                   \label{eq:26}
\end{eqnarray}
\noindent where
\begin{eqnarray}
\bar{T}(s_{i}, t_{p}) &=& \bar{T}_{C}(t_{p}) +
\bar{T}_{D}(s_{i}),                                   \nonumber \\
\bar{T}(s_{f}, t_{p}) &=& \bar{T}_{C}(t_{p}) +
\bar{T}_{D}(s_{f}),                                   \nonumber \\
\bar{T}(s_{i}, t_{q}) &=& \bar{T}_{C}(t_{q}) +
\bar{T}_{D}(s_{i}),                                   \nonumber \\
\bar{T}(s_{f}, t_{q}) &=& \bar{T}_{C}(t_{q}) +
\bar{T}_{D}(s_{f}).                                   \nonumber
\end{eqnarray}
\noindent $\bar{T}_{C} (t_p)$ and $\bar{T}_{C} (t_q)$ are defined by
Eq.\ (\ref{eq:23a}), and $\bar{T}_{D} (s_i)$ and $\bar{T}_{D} (s_f)$
are defined by Eq.\ (\ref{eq:23b}). The internal bremsstrahlung
amplitude corresponding to Fig.~3(e) can be written as
\begin{eqnarray}
\bar{M}^{I(D)}_{\mu} &=& \sum_{l} (Q_{A} + Q_{B}) \Gamma^{ADB}_{l}
\frac{i}{(q_{f}+p_{f})^{2}-(m^{D}_{l})^{2} + i\epsilon} [-
i(q_{i}+p_{i}+q_{f}+p_{f}+K)_{\mu}] \nonumber \\
&\times&  \frac{i}{(q_{i}+p_{i})^{2}-(m^{D}_{l})^{2} + i \epsilon}
\Gamma^{ADB}_{l}.
                                                      \label{eq:27}
\end{eqnarray}
\noindent Applying the radiation decomposition identity of Brodsky
and Brown to
split the amplitude $\bar{M}^{I(D)}_{\mu}$, we obtain
\begin{mathletters}
\begin {eqnarray}
\bar{M}^{(I)D}_{\mu} &=& Q_{A} \bar{T}_{D}(s_{f})
\frac{(q_{i}+p_{i})_{\mu}}{(q_{i}+p_{i})\cdot K} - Q_{A}
\frac{(q_{f}+p_{f})_{\mu}}{(q_{f}+p_{f})\cdot K}
\bar{T}_{D}(s_{i})                                    \nonumber \\
&+& Q_{B} \bar{T}_{D}(s_{f})
\frac{q_{i}+p_{i})_{\mu}}{(q_{i}+p_{i})\cdot K} - Q_{B}
\frac{(q_{f}+p_{f})_{\mu}}{(q_{f}+p_{f})\cdot K}
\bar{T}_{D}(s_{i})                                    \label{eq:28a}
\end{eqnarray}
\noindent This can be expressed directly in terms of the T-matrices
defined
above plus an exchange term:

\begin{eqnarray}
\bar{M}^{(I)D}_{\mu} &=& Q_{A} [\bar{T}_{D}(s_{f}) +
\bar{T}_{C}(t_{p}) - \bar{T}_{C}(t_{p})]
\frac{(q_{i}+p_{i})_{\mu}}{(q_{i}+p_{i})\cdot K}
                                                        \nonumber \\
&-& Q_{A} \frac{(q_{f}+p_{f})_{\mu}}{(q_{f}+p_{f})\cdot K}
[\bar{T}_{D}(s_{i}) + \bar{T}_{C}(t_{p}) - \bar{T}_{C}(t_{p})]
                                                       \nonumber \\
&+& Q_{B} [\bar{T}_{D}(s_{f}) + \bar{T}_{C}(t_{q}) -
\bar{T}_{C}(t_{q})] \frac{(q_{i}+p_{i})_{\mu}}{(q_{i}+p_{i})\cdot K}
                                                     \label{eq:28b} \\
&-& Q_{B} \frac{(q_{f}+p_{f})_{\mu}}{(q_{f}+p_{f})\cdot K}
[\bar{T}_{D}(s_{i}) + \bar{T}_{C}(t_{q}) - \bar{T}_{C}(t_{q})]
                                                        \nonumber \\
\medskip
&=& Q_{A} \bar{T}(s_{f}, t_{p})
\frac{(q_{i}+p_{i})_{\mu}}{(q_{i}+p_{i})\cdot K} - Q_{A}
\frac{(q_{f}+p_{f})_{\mu}}{(q_{f}+p_{f})\cdot K} \bar{T}(s_{i},
t_{p})                                                   \nonumber \\
&+& Q_{B} \bar{T}(s_{f}, t_{q})
\frac{(q_{i}+p_{i})_{\mu}}{(q_{i}+p_{i})\cdot K} - Q_{B}
\frac{(q_{f}+p_{f})_{\mu}}{(q_{f}+p_{f})\cdot K} \bar{T}(s_{i},
t_{q})                                                   \nonumber \\
&+& \bar{M}^{x}_{\mu},
                                                        \label{eq:28c}
\end{eqnarray}
\end{mathletters}
\noindent where
\begin{eqnarray}
\bar{M}^{x}_{\mu} &=& -Q_{A}  \bar{T}_{C}(t_{p})
\frac{(q_{i}+p_{i})_{\mu}}{(q_{i}+p_{i})\cdot K} + Q_{A}
\frac{(q_{f}+p_{f})_{\mu}}{(q_{f}+p_{f})\cdot K}
\bar{T}_{C}(t_{p})                                      \nonumber \\
&-& Q_{B} \bar{T}_{C}(t_{q})
\frac{(q_{i}+p_{i})_{\mu}}{(q_{i}+p_{i})\cdot K} - Q_{B}
\frac{(q_{f}+p_{f})_{\mu}}{(q_{f}+p_{f})\cdot K} \bar{T}_{C}(t_{q}).
                                                        \label{eq:29}
\end{eqnarray}
\noindent Neglecting $\bar{M}^{x}_{\mu}\,\,(\bar{M}^{x}_{\mu}\cdot
\epsilon^{\mu} = 0$ because the T-channel contribution $\equiv$ 0),
the
expression for $\bar{M}^{I(D)}$ in terms of the four quasi external
amplitudes
becomes
\begin{eqnarray}
\bar{M}^{I(D)}_{\mu} &=& -Q_{A} \bar{T}(s_{f},
t_{p})\frac{(q_{i}+p_{i})_{\mu}}{(q_{i}+p_{i})\cdot K} - Q_{A}
\frac{(q_{f}+p_{f})_{\mu}}{(q_{f}+p_{f})\cdot K}
\bar{T}(s_{i}, t_{p})                                  \nonumber \\
&+& Q_{B} \bar{T}(s_{f}, t_{q})
\frac{((q_{i}+p_{i})_{\mu}}{(q_{i}+p_{i})\cdot K} - Q_{B}
\frac{(q_{f}+p_{f})_{\mu}}{(q_{f}+p_{f})\cdot K} \bar{T}(s_{i},
t_{q}).                                                \label{eq:30}
\end{eqnarray}

We emphasize here that the expression for $\bar{M}^{I(D)}_{\mu}$
given by Eq.\ (\ref{eq:30}) is very general. That is, neglecting
$\bar{M}^{x}_{\mu}$ can be justified
on general grounds. To see this, consider

\[
A(q_{i}^{\mu}) + B(p_{i}^{\mu}) \longrightarrow A'(q_{f}^{\mu}) +
B'(p_{f}^{\mu}) + \gamma(K^{\mu}).                       \nonumber
\]
\noindent We assume that particles A and B have charges Q$_{A}$ and
Q$_B$,
respectively, while particles A$'$ and B$'$ have charges
Q$'_A$
and Q$'_B$, respectively. In this case, the amplitude
$\bar{M}^{E(CD)}_{\mu}$ given by Eq.\ (\ref{eq:26}) becomes
$\bar{M}^{E(CD)}_{\mu}$,

\begin{eqnarray}
\tilde{M}^{E(CD)}_{\mu}
&=& Q'_{A} \frac{q_{f\mu}}{q_{f}\cdot K} \bar{T}(s_{i}, t_{p}) -
Q_{A} \frac{q_{i\mu}}{q_{i}\cdot K} \bar{T}(s_{f}, t_{p}) \nonumber
\\
&+& Q'_{B} \frac{p_{f\mu}}{p_{f}\cdot K} \bar{T}(s_{i}, t_{q}) -
Q_{B} \frac{p_{i\mu}}{p_{i}\cdot K} \bar{T}(s_{f}, t_{q})
                                                        \label{eq:31}
\end{eqnarray}
\noindent while the amplitude $\bar{M}^{I}_{\mu}$ given by
Eq.\ (\ref{eq:28c}) becomes $\bar{M}^{I(D)}_{\mu}$,

\begin {eqnarray}
\tilde{M}^{I(D)}_{\mu} &=& Q_{A}  \bar{T}(s_{f}, t_{p})
\frac{(q_{i}+p_{i})_{\mu}}{(q_{i}+p_{i})\cdot K} - Q'_{A}
\frac{(q_{f}+p_{f})_{\mu}}{(q_{f}+p_{f})\cdot K}\bar{T}(s_{i},
t_{p})                                                 \nonumber \\
&+& Q_{B}  \bar{T}(s_{f}, t_{q})
\frac{(q_{i}+p_{i})_{\mu}}{(q_{i}+p_{i})\cdot K} - Q'_{B}
\frac{(q_{f}+p_{f})_{\mu}}{(q_{f}+p_{f})\cdot K} \bar{T}(s_{i},
t_{q})
                                                        \nonumber \\
&+& \tilde{M}^{x}_{\mu},
                                                       \label{eq:32}
\end{eqnarray}
\noindent where
\begin{eqnarray}
\tilde{M}^{x}_{\mu} &=& -Q_{A}  \bar{T}_{C}(t_{p})
\frac{(q_{i}+p_{i})_{\mu}}{(q_{i}+p_{i})\cdot K} + Q'_{A}
\frac{(q_{f}+p_{f})_{\mu}}{(q_{f}+p_{f})\cdot K}
\bar{T}_{C}(t_{p})                                     \nonumber \\
&-& Q_{B}  \bar{T}(t_{q})
\frac{(q_{i}+p_{i})_{\mu}}{(q_{i}+p_{i})\cdot K} - Q'_{B}
\frac{(q_{f}+p_{f})_{\mu}}{(q_{f}+p_{f})\cdot K} \bar{T}_{C}(t_{q}).
                                                      \label{eq:33}
\end{eqnarray}
\noindent Obviously, the amplitude

\begin{equation}
\tilde{M}^{EI}_{\mu} = \tilde{M}^{CD}_{\mu} +
\tilde{M}^{I(D)}_{\mu}
                                                     \label{eq:34}
\end{equation}
\noindent is not gauge invariant, since
\begin{eqnarray}
\tilde{M}^{EI}_{\mu} K^{\mu}
&=& \tilde{M}^{x}_{\mu} K^{\mu} \nonumber \\
&=& - Q_{A}\bar{T}_{C}(t_{p}) + Q'_{A}\bar{T}_{C}(t_{p}) -
Q_{B}\bar{T}_{C}(t_{q}) + Q'_{B}\bar{T}_{C}(t_{q}) \neq 0
                                                     \label{eq:35}
\end{eqnarray}
\noindent if Q$_A \neq$ Q$' _A$ and Q$_B \neq$ Q$' _B$.
Therefore, we
must construct an additional gauge term by imposing the condition
that the
total amplitude must be gauge invariant. Let $\tilde{M}_{\mu}$ be the
total
amplitude which is the sum of $\tilde{M}^{EI}_{\mu}$ and an
additional gauge
term $\tilde{M}^{G}_{\mu}$,
\begin{equation}
\tilde{M}_{\mu} = \tilde{M}^{EI}_{\mu} + \tilde{M}^{G}_{\mu}.
                                                      \label{eq:36}
\end{equation}
\noindent The gauge invariant condition demands that
\begin{eqnarray}
\tilde{M}_{\mu}K^{\mu} &=& \tilde{M}^{EI}_{\mu}K^{\mu} +
\tilde{M}^{G}_{\mu}K^{\mu} \nonumber \\
&=& \tilde{M}^{x}_{\mu}K^{\mu} + \tilde{M}^{G}_{\mu}K^{\mu} =0.
                                                     \label{eq:37}
\end{eqnarray}
\noindent It is clear that we may choose
\begin{equation}
\tilde{M}^{G}_{\mu} = -\tilde{M}^{x}_{\mu},
                                                    \label{eq:38}
\end{equation}
\noindent so that the term $\tilde{M}^{x}_{\mu}$ in
Eq.\ (\ref{eq:32}) is completely
canceled by the additional gauge term $\tilde{M}^{G}_{\mu}$. Hence,
we can in
general ignore the term $\tilde{M}^{x}_{\mu}$ in
Eq.\ (\ref{eq:32}), and therefore in
the special case of Q$_A$=Q$' _A$ and Q$_B$=Q$' _B$
described by Eq.\ (\ref{eq:28c}).

Combining the external amplitude of Eq.\ (\ref{eq:26}) and
the quasi external amplitudes
of Eq.\ (\ref{eq:30}), we obtain the total amplitude
$\bar{M}^{TsTts}_{\mu}$,
\begin{eqnarray}
\bar{M}^{TsTts}_{\mu} &=& \bar{M}^{E(CD)}_{\mu} +
\bar{M}^{I(D)}_{\mu} \nonumber \\
&=& Q_{A} [ \frac{q_{f\mu}}{q_{f}\cdot K} -
\frac{(q_{f}+p_{f})_{\mu}}{(q_{f}+p_{f})\cdot K} ] \bar{T}(s_{i},
t_{p}) \nonumber \\
&-& Q_{A} T(s_{f}, t_{p}) [ \frac{q_{i \mu}}{q_{i}\cdot K} -
\frac{(q_{i} + p_{i})_{\mu}}{(q_{i}+p_{i})\cdot K} ] \nonumber \\
&+& Q_{B} [ \frac{p_{f \mu}}{p_{f}\cdot K} -
\frac{(q_{f}+p_{f})_{\mu}}{(q_{f}+p_{f})\cdot K} ] \bar{T}(s_{i},
t_{q})
                                                  \label{eq:39}\\
&-& Q_{B} T(s_{f}, t_{q}) [ \frac{p_{i \mu}}{p_{i}\cdot K} -
\frac{(q_{i} + p_{i})_{\mu}}{(q_{i}+p_{i})\cdot K} ] \nonumber
\end{eqnarray}
\noindent It is easy to show that $\bar{M}^{TsTts}_{\mu}$ is gauge
invariant; that is,
\begin{equation}
\bar{M}^{TsTts}_{\mu}K^{\mu} = 0.
                                                   \label{eq:40}
\end{equation}
\noindent Here, we have used ``TsTts'' to identify the amplitude
given by Eq.~(\ref{eq:39}), because the amplitude can be
classified as the two-s-two-t special (TsTts) amplitude [22].

\subsection{Photon emission from the tree diagrams given by
Fig.~1(c)}

Using the elastic scattering diagrams given by Fig.~1(c) as source
graphs to
generate bremsstrahlung diagrams, we obtain Fig.~4. Figs.~4(a)-4(d)
represent the
external emission diagrams and Fig.~4(e) represents the internal
emission
diagrams. The external bremsstrahlung amplitude corresponding to
Figs.~4(a)-4(d)
has the form [23]
\begin{eqnarray}
\bar{M}^{E(CF)}_{\mu} &=& Q_{A} \frac{q_{f\mu}}{q_{f}\cdot
K}\bar{T}(u_{1}, t_{p}) - Q_{A} \frac{q_{i\mu}}{q_{i}\cdot K}
\bar{T}(u_{2}, t_{p}) \nonumber \\
&+& Q_{B}  \frac{p_{f\mu}}{p_{f}\cdot K}\bar{T}(u_{2}, t_{q}) -
Q_{B} \frac{p_{i\mu}}{p_{i}\cdot K} \bar{T}(u_{1}, t_{q})
                                                      \label{eq:41}
\end{eqnarray}
\noindent where
\begin{eqnarray}
\bar{T}(u_{1}, t_{p}) &=& \bar{T}_{C}(t_{p}) + \bar{T}_{F}(u_{1}),
                                                      \nonumber \\
\bar{T}(u_{2}, t_{p}) &=& \bar{T}_{C}(t_{p}) + \bar{T}_{F}(u_{2}),
                                                      \nonumber \\
\bar{T}(u_{2}, t_{q}) &=& \bar{T}_{C}(t_{q}) + \bar{T}_{F}(u_{2}),
                                                      \nonumber \\
\bar{T}(u_{1}, t_{q}) &=& \bar{T}_{C}(t_{q}) + \bar{T}_{F}(u_{1}).
                                                      \nonumber
\end{eqnarray}
$\bar{T}_{C}(t_{p})$ and $\bar{T}_{C}(t_{q})$ are defined by
Eq.\ (\ref{eq:23a}),
$\bar{T}_{F}(u_{1})$ and $\bar{T}_{F}(u_{2})$ are defined by
Eq.\ (\ref{eq:25}), and
u$_1$ and u$_2$ are defined by Eqs. (\ref{eq:11}) and (\ref{eq:14}),
respectively.  The internal bremsstrahlung amplitude corresponding
to Fig.~4(e) can be  written as
\begin{eqnarray}
\bar{M}^{I(F)}_{\mu} &=& \sum_{j} (Q_{A}-Q_{B})\Gamma^{AFB}_{j}
\frac{i}{(q_{i}-p_{f}-K)^{2}-(m^{F}_{j})^{2} + i\epsilon}[-i(q_{i}-
p_{f}+q_{i}-p_{f}-K)_{\mu}]                            \nonumber \\
&\times& \frac{i}{(q_{i}-p_{f})^{2} - (m^{F}_{j})^{2} + i\epsilon}
\Gamma^{AFB}_{j}
                                                       \label{eq:42}
\end{eqnarray}
\noindent which can be decomposed by using the Brodsky-Brown identity
as was
done with  (3.2). The decomposed amplitude
\begin{mathletters}
\begin{eqnarray}
\bar{M}^{I(F)}_{\mu} &=& Q_{A}\bar{T}_{F}(u_{2}) \frac{(q_{i}-
p_{f})_{\mu}}{(q_{i}-p_{f})\cdot K} - Q_{A} \frac{(p_{i}-
q_{f})_{\mu}}{(p_{i}-q_{f})\cdot K}
\bar{T}_{F}(u_{1})                                      \nonumber \\
&+& Q_{B}\bar{T}_{F}(u_{1}) \frac{(p_{i}-q_{f})_{\mu}}{(p_{i}-
q_{f})\cdot K} - Q_{B} \frac{(q_{i}-p_{f})_{\mu}}{(q_{i}-p_{f})\cdot
K} \bar{T}_{F}(u_{2})                                \label{eq:43a}
\end{eqnarray}
\noindent can be written as
\begin{eqnarray}
\bar{M}^{I(F)}_{\mu} &=& Q_{A}[\bar{T}_{F}(u_{2}) +
\bar{T}_{C}(t_{p}) - \bar{T}_{C}(t_{p})] \frac{(q_{i}-
p_{f})_{\mu}}{(q_{i}-p_{f})\cdot K}                     \nonumber \\
&-& Q_{A}\frac{(p_{i}-q_{f})_{\mu}}{(p_{i}-q_{f})\cdot
K}[\bar{T}_{F}(u_{1}) + \bar{T}_{C}(t_{p}) - \bar{T}_{C}(t_{p})]
                                                        \nonumber \\
&+& Q_{B}[\bar{T}_{F}(u_{1}) + \bar{T}_{C}(t_{q}) -
\bar{T}_{C}(t_{q})] \frac{(p_{i}-q_{f})_{\mu}}{(p_{i}-q_{f})\cdot K}
                                                   \label{eq:43b} \\
&-& Q_{B}\frac{(q_{i}-p_{f})_{\mu}}{(q_{i}-p_{f})\cdot
K}[\bar{T}_{F}(u_{2}) + \bar{T}_{C}(t_{q)} - \bar{T}_{C}(t_{q})]
                                                        \nonumber \\
\medskip
&=& Q_{A} \bar{T}(u_{2}, t_{p}) \frac{(q_{i}-p_{f})_{\mu}}{(q_{i}-
p_{f})\cdot K} - Q_{A} \frac{(p_{i}-q_{f})_{\mu}}{(p_{i}-q_{f})\cdot
K}\bar{T}(u_{1}, t_{p})                                 \nonumber \\
&+& Q_{B} \bar{T}(u_{1}, t_{q}) \frac{(p_{i}-q_{f})_{\mu}}{(p_{i}-
q_{f})\cdot K} - Q_{B} \frac{(q_{i}-p_{f})_{\mu}}{(q_{i}-p_{f})\cdot
K}\bar{T}(u_{2}, t_{q})                                  \nonumber \\
&+& \bar{M}^{Y}_{\mu},
                                                       \label{eq:43c}
\end{eqnarray}
\end{mathletters}
\noindent where
\begin{eqnarray}
\bar{M}^{Y}_{\mu} &=& Q_{A}\bar{T}_{C}(t_{p}) \frac{(q_{i}-
p_{f})_{\mu}}{(q_{i}-p_{f})\cdot K} + Q_{A} \frac{(p_{i}-
q_{f})_{\mu}}{(p_{i}-q_{f})\cdot K}
\bar{T}_{C}(t_{p})                                      \nonumber \\
&-& Q_{B}\bar{T}_{C}(t_{q}) \frac{(p_{i}-q_{f})_{\mu}}{(p_{i}-
q_{f})\cdot K} + Q_{A} \frac{(q_{i}-p_{f})_{\mu}}{(q_{i}-p_{f})\cdot
K} \bar{T}_{C}(t_{q}).
                                                        \label{eq:44}
\end{eqnarray}
\noindent Again, we can apply the same reasoning given in last
section, III.~A, to neglect the term
$\bar{M}^{Y}_{\mu}$
($\equiv 0$ in this case). Hence, we obtain the four quasi external
amplitudes
\begin{eqnarray}
\bar{M}^{I(F)}_{\mu} &=& Q_{A}\bar{T}(u_{2}, t_{p}) \frac{(q_{i}-
p_{f})_{\mu}}{(q_{i}-p_{f})\cdot K} - Q_{A} \frac{(p_{i}-
q_{f})_{\mu}}{(p_{i}-q_{f})\cdot K}
\bar{T}(u_{1}, t_{p})                                 \nonumber \\
&+& Q_{B}\bar{T}(u_{1}, t_{q}) \frac{(p_{i}-q_{f})_{\mu}}{(p_{i}-
q_{f})\cdot K} - Q_{B} \frac{(q_{i}-p_{f})_{\mu}}{(q_{i}-p_{f})\cdot
K} \bar{T}(u_{2}, t_{q}).
                                                       \label{eq:45}
\end{eqnarray}
\noindent The total amplitude $\bar{M}^{TuTts}_{\mu}$ is therefore
the sum of
$\bar{M}^{E(CF)}_{\mu}$ and $\bar{M}^{I(F)}_{\mu}$ [given by
Eq.\ (\ref{eq:20})]:
\begin{eqnarray}
\bar{M}^{TuTts}_{\mu} &=& \bar{M}^{E(CF)}_{\mu} +
\bar{M}^{I(F)}_{\mu}                                   \nonumber \\
&=& Q_{A} [ \frac{q_{f\mu}}{q_{f}\cdot K} - \frac{(p_{i}-
q_{f})_{\mu}}{(p_{i}-q_{f})\cdot K} ] \bar{T}(u_{1}, t_{p})
                                                       \nonumber \\
&-& Q_{A} T(u_{2}, t_{p}) [ \frac{q_{i\mu}}{q_{i}\cdot K} -
\frac{(q_{i}-p_{f})_{\mu}}{(q_{i}-p_{f})\cdot K} ]      \nonumber \\
&+& Q_{B} [ \frac{p_{f\mu}}{p_{f}\cdot K} - \frac{(q_{i}-
p_{f})_{\mu}}{(q_{i}-p_{f})\cdot K} ] \bar{T}(u_{2}, t_{q})
                                                    \label{eq:46} \\
&-& Q_{B} T(u_{1}, t_{q}) [ \frac{p_{i\mu}}{p_{i}\cdot K} -
\frac{(p_{i}-q_{f})_{\mu}}{(p_{i}-q_{f})\cdot K} ]   \nonumber
\end{eqnarray}
\noindent Obviously, the amplitude $\bar{M}^{TuTts}_{\mu}$ is gauge
invariant; that is,
\begin{equation}
\bar{M}^{TuTts}_{\mu} K^{\mu} = 0.
                                                     \label{eq:47}
\end{equation}
\noindent We have classified this amplitude as the two-u-two-t
special (TuTts)
amplitude [24].\

It should be pointed out that if we change p$^{\mu}_{i}$ to
$-$p$^{\mu}_{f}$, p$^{\mu}_{f}$ to $-$p$^{\mu}_{i}$ and Q$_B$ to
$-$Q$_B$, then
the amplitude $\bar{M}^{TuTts}_{\mu}$ becomes the amplitude
$\bar{M}^{TsTts}_{\mu}$:
\begin{equation}
\bar{M}^{TuTts}_{\mu} \frac{Q_{B}\longrightarrow -
Q_{B}}{p^{\mu}_{i}\longleftrightarrow -p^{\mu}_{f}}\!\!\!\rightarrow
\bar{M}^{TsTts}_{\mu}.
                                                   \label{eq:48}
\end{equation}
\medskip
\noindent The reverse is also true. This interchange equivalence is
expected
from a close examination of Fig.~1(c) and Fig.~1(b).\\
\section{SOFT-PHOTON AMPLITUDE}\label{sec:4}
\medskip
\indent If the elastic scattering diagram given by Fig.~1(a) is used
as the
source graph to generate a set of bremsstrahlung diagrams, we obtain
Fig.~2.
Figs.~2(a)-2(d) are the external emission diagrams and Fig.~2(e) is the
internal
emission diagram. $T_{a}, T_{b}, T_{c}$ and $T_d$ in these diagrams
represent
the half-off-shell T-matrices. It is well-known that there is no
general method
which can be used to determine the exact internal amplitude without
introducing
dynamical models. It is also true that it is difficult to calculate
all internal terms derived from a given model without introducing some
approximations. This is why various soft-photon amplitudes,
approximate
amplitudes consistent with the soft-photon theorem, have been
constructed and
applied to describe many different nuclear bremsstrahlung processes.
In the
past, the utility of these amplitudes was determined only by
comparison with
experimental measurements. Recently, however, there has been some
effort to
determine the range of validity of various soft-photon amplitudes
theoretically
without comparing with experimental data. Here, we investigate
methods for
selecting optimal independent Lorentz invariants to parametrize the
T-matrices
($T_{a}, T_{b}, T_{c}, T_d$) in the soft-photon amplitudes. We show
that the
question of validity of a given soft-photon approximation is directly
related
to the choice of independent Lorentz invariants. Four different
soft-photon
amplitudes are derived using two different procedures: the standard
Low
procedure and our modified Low procedure. The first two amplitudes
are derived
in subsections (A) and (B) and the last two amplitudes, which are
more general,
are derived in subsections (C) and (D).

(A) Below, we review the procedure for deriving the first
of two
Low's soft-photon amplitudes. The independent Lorentz invariants are
$s_{x} (x
= i, f),\,\, t_{y} (y= p, q)$ and $\Delta_{z} (z = a, b, c, d)$.
(These
invariants were defined in section II.) In other words, the four
half-off-shell
T-matrices are chosen to be
\begin{eqnarray}
T_{a} &=& T(s_{i}, t_{p}, \Delta_{a}),             \nonumber \\
T_{b} &=& T(s_{f}, t_{p}, \Delta_{b}),
                                                  \label{eq:49}\\
T_{c} &=& T(s_{i}, t_{q}, \Delta_{c}),                \nonumber
\end{eqnarray}
\noindent and
\[
T_{d} = T(s_{f}, t_{q}, \Delta_{d}),
\]
\noindent In terms of the above T-matrices, the external amplitude
can be
written in the familiar form

\begin{eqnarray}
M^{E}_{\mu}(s, t, \Delta) &=& Q_{A}\frac{q_{f\mu}}{q_{f}\cdot K}
T(s_{i}, t_{p},\Delta_{a}) - Q_{A} T(s_{f}, t_{p},
\Delta_{b})\frac{q_{i\mu}}{q_{i}\cdot K}
                                                       \nonumber \\
&+& Q_{B}\frac{p_{f\mu}}{p_{f}\cdot K} T(s_{i}, t_{p}, \Delta_{c}) -
Q_{B}T(s_{f}, t_{p}, \Delta_{d})\frac{p_{i\mu}}{p_{i}\cdot K}.
                                                        \label{eq:50}
\end{eqnarray}
\noindent Following Low, we introduce the average values of s and t:
\begin{eqnarray}
\bar{s} &=& \frac{1}{2}(s_{i} + s_{f})                   \nonumber \\
\bar{t} &=& \frac{1}{2}(t_{p} + t_{q}).                 \label{eq:51}
\end{eqnarray}
\noindent It is then easily demonstrated that
\begin{eqnarray}
s_{i} = \bar{s} + (q_{i} + p_{i})\cdot K &=& \bar{s} + (q_{f} +
p_{f})\cdot K,                                            \nonumber \\
s_{f} = \bar{s} - (q_{i} + p_{i})\cdot K &=& \bar{s} - (q_{f} +
p_{f})\cdot K,                                          \nonumber \\
t_{p} = \bar{t} - (q_{i} - q_{f})\cdot K &=& \bar{t} + (p_{i} -
p_{f})\cdot K,
                                                         \label{eq:52}
\end{eqnarray}
\noindent and
\[
t_{q} = \bar{t} + (q_{i} - q_{f})\cdot K = \bar{t} - (p_{i} -
p_{f})\cdot K.
\]
\noindent If all half-off-shell T-matrices are expanded about
[$\bar{s},
\bar{t}, \Delta = (mass)^{2}$], then we obtain
\begin{eqnarray}
M^{E}_{\mu}(s, t, \Delta) &=& Q_{A}\frac{q_{f\mu}}{q_{f}\cdot K}
[T(\bar{s}, \bar{t}) + \frac{\partial T(\bar{s}, \bar{t})}{\partial
\bar{s}}(q_{i} +
p_{i})\cdot K + \frac{\partial T(\bar{s}, \bar{t})}{\partial
\bar{t}}(p_{i} - p_{f})\cdot K + \frac{\partial T_{a}}{\partial
\Delta_{a}}2q_{f}\cdot K]
                                                       \nonumber \\
&-& Q_{A}[T(\bar{s}, \bar{t}) - \frac{\partial T(\bar{s},
\bar{t})}{\partial \bar{s}}(q_{i} + p_{i})\cdot K + \frac{\partial
T(\bar{s}, \bar{t})}{\partial
\bar{t}}(p_{i} - p_{f})\cdot K - \frac{\partial T_{b}}{\partial
\Delta_{b}}2q_{i}\cdot K]\frac{q_{i\mu}}{q_{i}\cdot K}
                                                       \nonumber \\
&+& Q_{B}\frac{p_{f\mu}}{p_{f}\cdot K} [T(\bar{s}, \bar{t}) +
\frac{\partial
T(\bar{s}, \bar{t})}{\partial \bar{s}}(q_{i} + p_{i})\cdot K -
\frac{\partial
T(\bar{s}, \bar{t})}{\partial \bar{t}}(p_{i} - p_{f})\cdot K +
\frac{\partial
T_{c}}{\partial \Delta_{c}}2p_{f}\cdot K]             \nonumber \\
&-& Q_{B}[T(\bar{s}, \bar{t}) - \frac{\partial T(\bar{s},
\bar{t})}{\partial
\bar{s}}(q_{i} + p_{i})\cdot K - \frac{\partial T(\bar{s},
\bar{t})}{\partial
\bar{t}}(p_{i} - p_{f})\cdot K - \frac{\partial T_{d}}{\partial
\Delta_{d}}2p_{i}\cdot K]\frac{p_{i\mu}}{p_{i}\cdot K}
                                                       \nonumber \\
&+& O(K),                                             \label{eq:53}
\end{eqnarray}
\noindent where $T(\bar{s}, \bar{t}) \equiv T(\bar{s}, \bar{t},
m^{2}_{A},
m^{2}_{B}, m^{2}_{A}, m^{2}_{B})$ is the elastic scattering
(on-shell) T-matrix
evaluated at $\bar{s}$ and $\bar{t}$. Now, we impose the gauge
invariant condition
\[
[M^{E}_{\mu}(s, t, \Delta) + M^{I}_{\mu}(s, t, \Delta)] K^{\mu} = 0,
\]
\noindent which gives
\begin{eqnarray}
M^{I}_{\mu} K^{\mu} &=& -2(Q_{A} + Q_{B})\frac{\partial T(\bar{s},
\bar{t})}{\partial \bar{s}}(q_{i} + p_{i})\cdot K -
2Q_{A}\frac{\partial
T_{a}}{\partial \Delta_{a}}q_{f}\cdot K - 2Q_{A}\frac{\partial
T_{b}}{\partial
\Delta_{b}}q_{i}\cdot K                                 \nonumber \\
&-& 2Q_{B}\frac{\partial T_{c}}{\partial \Delta_{c}}p_{f}\cdot K -
2Q_{B}\frac{\partial T_{d}}{\partial \Delta_{d}}p_{i}\cdot K.
                                                        \label{eq:54}
\end{eqnarray}
\noindent Hence, the leading term of $M^{I}_{\mu}(s, t, \Delta)$ has
the form:

\begin{eqnarray}
M^{I}_{\mu}(s, t, \Delta) &=& -2(Q_{A} + Q_{B})\frac{\partial
T(\bar{s},
\bar{t})}{\partial \bar{s}}(q_{i} + p_{i})_{\mu} -
2Q_{A}\frac{\partial
T_{a}}{\partial \Delta_{a}}q_{f\mu}                      \nonumber \\
&-& 2Q_{B}\frac{\partial T_{b}}{\partial \Delta_{b}}q_{i\mu} -
2Q_{A}\frac{\partial T_{c}}{\partial \Delta_{c}}p_{f\mu} -
2Q_{B}\frac{\partial T_{d}}{\partial \Delta_{d}}p_{i\mu}.
                                                        \label{eq:55}
\end{eqnarray}
\noindent Combining Eqs.\ (\ref{eq:53}) and (\ref{eq:55}), we obtain
the total  bremsstrahlung amplitude $M^{Low(st)}_{\mu}$

\begin{equation}
M^{Low(st)}_{\mu} = M^{E(st)}_{\mu} + M^{I(st)}_{\mu}
                                                       \label{eq:56}
\end{equation}
\noindent where $M^{E(st)}_{\mu}$ is on the on-shell part of the
external
amplitude which depends on $\bar{s}$ and $\bar{t}$,

\begin{eqnarray}
M^{E(st)}_{\mu} &=& [Q_{A}(\frac{q_{f\mu}}{q_{f}\cdot K} -
\frac{q_{i\mu}}{q_{i}\cdot K}) + Q_{B}(\frac{p_{f\mu}}{p_{f}\cdot K}
-\frac{p_{i\mu}}{p_{i}\cdot K})]
T(\bar{s}, \bar{t})                                   \nonumber \\
&+& [Q_{A}(\frac{q_{f\mu}}{q_{f}\cdot K} + \frac{q_{i\mu}}{q_{i}\cdot
K}) + Q_{B}(\frac{p_{f\mu}}{p_{f}\cdot K} +
\frac{p_{i\mu}}{p_{i}\cdot
K})] (q_{i} +
p_{i})\cdot K \frac{\partial T(\bar{s}, \bar{t})}{\partial \bar{s}}
                                                    \label{eq:57}\\
&+& [Q_{A}(\frac{q_{f\mu}}{q_{f}\cdot K} - \frac{q_{i\mu}}{q_{i}\cdot
K}) - Q_{B}(\frac{p_{f\mu}}{p_{f}\cdot K} -
\frac{p_{i\mu}}{p_{i}\cdot
K})] (p_{i} -
p_{f})\cdot K \frac{\partial T(\bar{s}, \bar{t})}{\partial \bar{t}},
                                                        \nonumber
\end{eqnarray}
\noindent and $M^{I(st)}_{\mu}$ is the on-shell part of the internal
amplitude
which depends on $\bar{s}$ and $\bar{t}$,

\begin{equation}
M^{I(st)}_{\mu} = -2 (Q_{A} + Q_{B})(q_{i} + p_{i})_{\mu}
\frac{\partial T(\bar{s}, \bar{t})}{\partial \bar{s}}.
                                                        \label{eq:58}
\end{equation}
\noindent It is clear the $M^{I(st)}_{\mu}\epsilon^{\mu}$ contributes
nothing to the bremsstrahlung cross section since ($q_{i} +
p_{i})_{\mu}\epsilon^{\mu}$
vanishes in the C.~M. system and in the Coulomb gauge.\\
\medskip
\noindent (B) A second Low soft-photon amplitude can be derived if
the
independent Lorentz invariants are chosen to be $u_{j}\, (j = 1, 2),
t_{y}\, (y = p, q)$ and $\Delta_{z}\, (z = a, b, c, d)$.  Here,
$u_{1}$
and $u_{2}$ are defined by Eqs. (\ref{eq:11}) and (\ref{eq:14}),
respectively.  With this choice, the four half-off-shell
T-matrices are parametrized in terms of $u_{j}, t_{y}$ and
$\Delta_{z}$ as
\begin{eqnarray}
T_{a} = T(u_{1}, t_{p}, \Delta_{a}),                 \nonumber \\
T_{b} = T(u_{2}, t_{p}, \Delta_{b}),                 \nonumber \\
T_{c} = T(u_{2}, t_{q}, \Delta_{c}),                 \label{eq:59}
\end{eqnarray}
\noindent and
\begin{equation}
T_{d} = T(u_{1}, t_{q}, \Delta_{d}).                 \nonumber
\end{equation}
\noindent The external amplitude has the form
\begin{eqnarray}
M^{E}_{\mu}(u, t, \Delta) &=& Q_{A}\frac{q_{f\mu}}{q_{f}\cdot K}
T(u_{1}, t_{p},\Delta_{a}) - Q_{A} T(u_{2}, t_{p},
\Delta_{b})\frac{q_{i\mu}}{q_{i}\cdot K}
                                                     \nonumber \\
&+& Q_{B}\frac{p_{f\mu}}{p_{f}\cdot K} T(u_{2}, t_{p},\Delta_{c}) -
Q_{B} T(u_{1}, t_{p}, \Delta_{d})\frac{p_{i\mu}}{p_{i}\cdot K}.
                                                    \label{eq:60}
\end{eqnarray}
\noindent Introducing the average u,
\begin{equation}
\bar{u} = \frac{1}{2}(u_{1} + u_{2}),                \label{eq:61}
\end{equation}
\noindent we have
\begin{mathletters}
\begin{equation}
u_{1} = \bar{u} - (p_{f} - q_{i})\cdot K = \bar{u} - (p_{i} -
q_{f})\cdot K .                                      \label{eq:62a}
\end{equation}
\noindent and
\begin{equation}
u_{2} = \bar{u} - (p_{f} - q_{i})\cdot K = \bar{u} + (p_{i} -
q_{f})\cdot K .                                      \label{eq:62b}
\end{equation}
\end{mathletters}
\noindent If we expand all half-off-shell T-matrices in Eq.\ (\ref{eq:60})
about
[$\bar{u}, \bar{t}, \Delta = (mass)^{2}$], we find
\begin{eqnarray}
M^{E}_{\mu}(u, t, \Delta) &=& Q_{A}\frac{q_{f\mu}}{q_{f}\cdot K}
[T(\bar{u}, \bar{t}) - \frac{\partial T(\bar{u}, \bar{t})}{\partial
\bar{u}}(p_{f} -
q_{i})\cdot K + \frac{\partial T(\bar{u}, \bar{t})}{\partial
\bar{t}}(p_{i} - p_{f})\cdot K + \frac{\partial T_{a}}{\partial
\Delta_{a}}2q_{f}\cdot K]
                                                        \nonumber \\
&-& Q_{A}[T(\bar{u}, \bar{t}) + \frac{\partial T(\bar{u},
\bar{t})}{\partial \bar{u}}(p_{f} -
q_{i})\cdot K + \frac{\partial T(\bar{u}, \bar{t})}{\partial
\bar{t}}(p_{i} - p_{f})\cdot K - \frac{\partial T_{b}}{\partial
\Delta_{b}}2q_{i}\cdot K]
\frac{q_{i\mu}}{q_{i}\cdot K}                           \nonumber \\
&+& Q_{B}\frac{p_{f\mu}}{p_{f}\cdot K} [T(\bar{u}, \bar{t}) +
\frac{\partial T(\bar{u}, \bar{t})}{\partial \bar{u}}(p_{f} -
q_{i})\cdot K - \frac{\partial T(\bar{u}, \bar{t})}{\partial
\bar{t}}(p_{i} - p_{f})\cdot K + \frac{\partial T_{c}}{\partial
\Delta_{c}}2p_{f}\cdot K]
                                                        \nonumber \\
&-& Q_{B}[T(\bar{u}, \bar{t}) - \frac{\partial T(\bar{u},
\bar{t})}{\partial \bar{u}}(p_{f} -
q_{i})\cdot K + \frac{\partial T(\bar{u}, \bar{t})}{\partial
\bar{t}}(p_{i} - p_{f})\cdot K - \frac{\partial T_{d}}{\partial
\Delta_{d}}2p_{i}\cdot K]
\frac{p_{i\mu}}{p_{i}\cdot K}                            \nonumber \\
&+& 0(K).                                               \label{eq:63}
\end{eqnarray}
\noindent Here, $T(\bar{u}, \bar{t}) \equiv T(\bar{u}, \bar{t},
m^{2}_{A}, m^{2}_{B}, m^{2}_{A}, m^{2}_{B})$ is the elastic
(on-shell)
T-matrix evaluated at
$\bar{u}$ and $\bar{t}$.  To obtain the leading term of the internal
amplitude $M^{I}_{\mu}(u, t, \Delta)$, we again impose the gauge
invariant condition
\begin{equation}
[M^{E}_{\mu}(u, t, \Delta) + M^{I}_{\mu}(u, t, \Delta)] K^{\mu} = 0
                                                        \label{eq:64}
\end{equation}
\noindent from which we obtain
\begin{eqnarray}
M^{I}_{\mu}K^{\mu} &=& -2(Q_{A} - Q_{B})\frac{\partial T(\bar{u},
\bar{t})}{\partial \bar{u}}(p_{f} -
q_{i})\cdot K - 2Q_{A}\frac{\partial T_{a}}{\partial
\Delta_{a}}q_{f}\cdot K - 2Q_{A}\frac{\partial T_{b}}{\partial
\Delta_{b}}q_{i}\cdot K                                \nonumber \\
&-& 2Q_{B}\frac{\partial T_{c}}{\partial \Delta_{c}}p_{f}\cdot K -
2Q_{B}\frac{\partial T_{d}}{\partial \Delta_{d}}p_{i}\cdot K.
                                                      \label{eq:65}
\end{eqnarray}
\noindent Eq.\ (\ref{eq:65}) gives
\begin{eqnarray}
M^{I}_{\mu}(u, t, \Delta) &=& 2(Q_{A} - Q_{B})\frac{\partial
T(\bar{u}, \bar{t})}{\partial \bar{u}}(p_{f} -
q_{i})_{\mu} - 2Q_{A}\frac{\partial T_{a}}{\partial
\Delta_{a}}q_{f\mu}                                    \nonumber \\
&-& 2Q_{A}\frac{\partial T_{b}}{\partial \Delta_{b}}q_{i\mu} -
2Q_{B}\frac{\partial T_{c}}{\partial \Delta_{c}}p_{f\mu} -
2Q_{b}\frac{\partial T_{d}}{\partial \Delta_{d}}p_{i\mu}.
                                                       \label{eq:66}
\end{eqnarray}
\noindent The total bremsstrahlung amplitude $M^{Low(ut)}_{\mu}$ is
the sum of Eq.\ (\ref{eq:63}) and Eq.\ (\ref{eq:66}):
\begin{equation}
M^{Low(st)}_{\mu} = M^{E(ut)}_{\mu} + M^{I(ut)}_{\mu}   \label{eq:67}
\end{equation}
\noindent where
\begin{eqnarray}
M^{E(ut)}_{\mu} &=& [Q_{A}(\frac{q_{f\mu}}{q_{f}\cdot K} -
\frac{q_{i\mu}}{q_{i}\cdot K}) + Q_{B}(\frac{p_{f\mu}}{p_{f}\cdot K}
-
\frac{p_{i\mu}}{p_{i}\cdot K})]
T(\bar{u}, \bar{t})                                \label{eq:68}\\
&+& [-Q_{A}(\frac{q_{f\mu}}{q_{f}\cdot K} +
\frac{q_{i\mu}}{q_{i}\cdot
K}) + Q_{B}(\frac{p_{f\mu}}{p_{f}\cdot K}
+ \frac{p_{i\mu}}{p_{i}\cdot K})] (p_{f} - q_{i})\cdot K
\frac{\partial T(\bar{u}, \bar{t})}{\partial \bar{u}}
                                                      \nonumber \\
&+& [Q_{A}(\frac{q_{f\mu}}{q_{f}\cdot K} - \frac{q_{i\mu}}{q_{i}\cdot
K}) - Q_{B}(\frac{p_{f\mu}}{p_{f}\cdot K}
- \frac{p_{i\mu}}{p_{i}\cdot K})] (p_{i} - p_{f})\cdot K
\frac{\partial T(\bar{u}, \bar{t})}{\partial \bar{t}},
                                                       \nonumber
\end{eqnarray}
\noindent and
\begin{equation}
M^{I}_{\mu} = 2(Q_{A} - Q_{B})\frac{\partial T(\bar{u},
\bar{t})}{\partial \bar{u}}(p_{f} -
q_{i})_{\mu}.
                                                   \label{eq:69}
\end{equation}
\noindent Again, $M^{E(ut)}_{\mu}$ is the on-shell part of the
external amplitude which depends on $\bar{u}$ and $\bar{t}$ while
$M^{I(ut)}_{\mu}$ is the on-shell part of the internal amplitude
which depends on $\bar{u}$ and $\bar{t}$.  Unlike the internal
amplitude
$M^{I(st)}_{\mu}\epsilon^{\mu}$ [Eq.\ (\ref{eq:58})] which is identically
zero,
the internal amplitude $M^{I(ut)}_{\mu}\epsilon^{\mu}$ does not
vanish when $Q_{A} \neq Q_{B}$.  (It should be remembered that we
consider
specifically an s-channel reaction here.)  Thus, we see by this
simple example that different choices of
independent variables (Lorentz invariants) can lead to different
soft-photon amplitudes.  We shall discuss this further in section
V.

(C) As we have already mentioned, more general soft-photon
amplitudes
can be derived by using the modified Low procedure described in
section I.  In using this new procedure, the
construction of the internal amplitude is guided by the elastic
scattering and the bremsstrahlung processes at the tree level.
For example, if the A-B elastic scattering is dominated by the
one-particle
s-channel exchange diagrams, then the internal amplitude will
be determined by photon emissions from the s-channel exchange
particles.
(Since a t-channel exchange particle should be neutral, there is no
internal emission from it.)  In this case, we should choose a set of
independent Lorentz invariants which includes s and t.  On the other
hand, if the A-B elastic scattering is dominated by the one-particle
u-channel
exchange diagrams, then the internal emissions will come
from the u-channel exchange particles, and we should choose a set of
independent Lorentz invariants which includes u and t.  In this
subsection, a general bremsstrahlung amplitude for a process whose
elastic scattering is dominated by the diagrams shown in Fig.~1(b)
will
be derived.  [The derivation of another general bremsstrahlung
amplitude for a process whose elastic scattering is dominated by the
diagrams shown in Fig.~1(c), will be discussed in the next subsection
(D).]

Choosing the set of independent Lorentz invariants which includes
$s_{x} (x = i, f),\, t_{y} (y = p, q)$, and $\Delta_{z} (z = a, b, c,
d)$, the external emission amplitude $M^{E}_{\mu}(s, t, \Delta)$ is
identical to that given by Eq.\ (\ref{eq:50}).  Because we assume that the
elastic scattering depicted in Fig.~1(a) is dominated by
the diagrams shown in Fig.~1(b) and, likewise, that the
bremsstrahlung processes represented in Fig.~2
are dominated by the diagrams shown in
Fig.~3, we can write the internal emission amplitude in the form
\begin{eqnarray}
M^{I(D)}_{\mu}(s, t, \Delta) &=& Y_{a}\,T(s_{i}, t_{p}, \Delta_{a}) +
T(s_{f}, t_{p}, \Delta_{b})\,Y_{b}                       \nonumber \\
&+& Y_{c}\,T(s_{i}, t_{q}, \Delta_{c}) + T(s_{f}, t_{q},
\Delta_{d})\,Y_{d}
                                                        \label{eq:70}
\end{eqnarray}
\noindent where $Y_{z}(z = a, b, c, d)$ are electromagnetic factors
to
be specified.  To determine $Y_{z}$, we demand that $M^{I(D)}_{\mu}$
reduce
to the expression for $\bar{M}^{I(D)}_{\mu}$ given by
Eq.\ (\ref{eq:30}) when the
general diagram in Fig.~2(e) reduces to the tree approximation in
Fig.~3(e).
Since $T(s_{x}, t_{Y}, \Delta_{z})$ reduces to $\bar{T}(s_{x},
t_{Y})$ in
the
tree approximation in this special case, we find
\begin{eqnarray}
Y_{a} &=& -Q_{A} \frac{(q_{f} + p_{f})_{\mu}}{(q_{f} + p_{f})\cdot K}
                                                         \nonumber \\
Y_{b} &=& Q_{A} \frac{(q_{i} + p_{i})_{\mu}}{(q_{i} + p_{i})\cdot K}
                                                         \nonumber \\
Y_{c} &=& -Q_{B} \frac{(q_{f} + p_{f})_{\mu}}{(q_{f} + p_{f})\cdot K}
                                                        \label{eq:71}
\end{eqnarray}
\noindent and
\[
Y_{d} = Q_{B} \frac{(q_{i} + p_{i})_{\mu}}{(q_{i} + p_{i})\cdot K}.
\]
\noindent Now, combining $M^{E}_{\mu}(s, t, \Delta)$ given
by Eq.\ (\ref{eq:50}) with
$M^{I(D)}_{\mu}(s, t, \Delta)$ given by Eqs.\ (\ref{eq:70}) and
(\ref{eq:71}), we obtain
\begin{eqnarray}
M^{TsTt)}_{\mu} &=& Q_{A}[\frac{q_{f\mu}}{q_{f}\cdot K} -
\frac{(q_{f} + p_{f})_{\mu}}{(q_{f} + p_{f})\cdot K}]
T(s_{i}, t_{p}, \Delta_{a})                             \nonumber \\
&-& Q_{A}\,T(s_{f}, t_{p}, \Delta_{b}) [ \frac{q_{i\mu}}{q_{i}\cdot
K} -
\frac{(q_{i} + p_{i})_{\mu}}{(q_{i} + p_{i})\cdot K}]
                                                        \nonumber \\
&+& Q_{B}[\frac{p_{f\mu}}{p_{f}\cdot K} -
\frac{(q_{f} + p_{f})_{\mu}}{(q_{f} + p_{f})\cdot K}] T(s_{i}, t_{q},
\Delta_{c})
                                                        \nonumber \\
&-& Q_{B}\,T(s_{f}, t_{q}, \Delta_{d}) [ \frac{p_{i\mu}}{p_{i}\cdot
K} -
\frac{(q_{i} + p_{i})_{\mu}}{(q_{i} + p_{i})\cdot K}].
                                                       \label{eq:72}
\end{eqnarray}
\noindent Because $M^{TsTt}_{\mu}$ is already explicitly gauge
invariant,
\[
M^{TsTt}_{\mu}K^{\mu} = 0,
\]
\noindent no additional gauge term is needed. The amplitude
$M^{TsTt}_{\mu}$ is
an off-shell two-s-two-t (TsTt) amplitude derived for the A-B
bremsstrahlung
process when internal emission from the s-channel exchange particles
is dominant. To obtain an on-shell TsTt special amplitude
$M^{TsTt}_{\mu}$ which
is free of any derivative of the T- matrix with respect to s or t, we
expand
$T(s_{x}, t_{Y}, \Delta_{z})$ only about the on-shell point
(mass)$^2$ in
$\Delta_{z}$:

\begin{eqnarray}
T(s_{i}, t_{p}, \Delta_{a}) &=& T(s_{i}, t_{p}) + \frac{\partial
T(s_{i},
t_{p}, \Delta_{a})}{\partial \Delta_{a}}2q_{f}\cdot K,
                                                        \nonumber \\
T(s_{f}, t_{p}, \Delta_{b}) &=& T(s_{f}, t_{p}) - \frac{\partial
T(s_{f},
t_{p}, \Delta_{b})}{\partial \Delta_{b}}2q_{i}\cdot K,
                                                        \nonumber \\
T(s_{i}, t_{q}, \Delta_{a}) &=& T(s_{i}, t_{q}) + \frac{\partial
T(s_{i},
t_{q}, \Delta_{c})}{\partial \Delta_{c}}2p_{f}\cdot K,
                                                       \label{eq:73}
\end{eqnarray}
\noindent and
\[
T(s_{f}, t_{q}, \Delta_{d}) = T(s_{f}, t_{q}) - \frac{\partial
T(s_{f},
t_{q}, \Delta_{d})}{\partial \Delta_{d}}2p_{i}\cdot K,
\]
\noindent where
\begin{eqnarray}
T(s_{i}, t_{p}) &\equiv& T(s_{i}, t_{p}, m^{2}_{A}), \nonumber \\
T(s_{f}, t_{p}) &\equiv& T(s_{f}, t_{p}, m^{2}_{A}), \nonumber \\
T(s_{i}, t_{q}) &\equiv& T(s_{i}, t_{q}, m^{2}_{B}), \nonumber
\end{eqnarray}
\noindent and
\[
T(s_{f}, t_{q}) \equiv T(s_{f}, t_{q}, m^{2}_{B}).
\]
\noindent Inserting Eq.\ (\ref{eq:73}) into Eq.\ (\ref{eq:72}) gives
\begin{equation}
M^{TsTt}_{\mu} = M^{TsTts}_{\mu} + M^{off(st)}_{\mu},   \label{eq:74}
\end{equation}
\noindent where
\begin{eqnarray}
M^{TsTt}_{\mu} &=& Q_{A} [ \frac{q_{f\mu}}{q_{f}\cdot K} -
\frac{(q_{f} +
p_{f})_{\mu}}{(q_{f} + p_{f})\cdot K} ] T(s_{i}, t_{p})
                                                         \nonumber \\
&-& Q_{A} \ T(s_{f}, t_{p}) [ \frac{q_{i\mu}}{q_{i}\cdot K} -
\frac{(q_{i}
+ p_{i})_{\mu}}{(q_{i} + p_{i})\cdot K} ]                \nonumber \\
&+& Q_{B} [ \frac{p_{f\mu}}{p_{f}\cdot K} - \frac{(q_{f} +
p_{f})_{\mu}}{(q_{f} + p_{f})\cdot K} ] T(s_{i}, t_{q})
                                                        \label{eq:75}\\
&-& Q_{B} \ T(s_{f}, t_{q}) [ \frac{p_{i\mu}}{p_{i}\cdot K} -
\frac{(q_{i}
+ p_{i})_{\mu}}{(q_{i} + p_{i})\cdot K} ]                \nonumber
\end{eqnarray}
\noindent and $M^{off(st)}_{\mu}$ represents those terms involving
off-shell
derivatives of the T-matrix. The amplitude $M^{off(st)}_{\mu}$ is
neglected in
the soft-photon approximation. The soft-photon amplitude
$M^{TsTts}_{\mu}$ is
the on-shell TsTt special amplitude and is more general than the
amplitude
$M^{Low(st)}_{\mu}$ given by Eq.\ (\ref{eq:56}), the soft-photon amplitude
derived by
using Low's standard procedure. To see this point, let us rewrite
$M^{TsTts}_{\mu}$ into two parts, an external term
$M^{E(TsTt)}_{\mu}$ and an
internal term $M^{I(TsTt)}_{\mu}$:
\begin{eqnarray}
M^{E(TsTt)}_{\mu} &=& Q_{A}  \frac{q_{f\mu}}{q_{f}\cdot K}T(s_{i},
t_{p})
- Q_{A} \ T(s_{f}, t_{p}) \frac{q_{i\mu}}{q_{i}\cdot K}
                                                        \nonumber \\
&+& Q_{B}  \frac{p_{f\mu}}{p_{f}\cdot K}T(s_{i}, t_{q})
- Q_{B} \ T(s_{f}, t_{q}) \frac{p_{i\mu}}{p_{i}\cdot K}
                                                        \label{eq:76}
\end{eqnarray}
\noindent and
\begin{eqnarray}
M^{I(TsTt)}_{\mu} &=& -\{Q_{A}[T(s_{i}, t_{p}) - T(s_{f}, t_{p})] +
Q_{B}[T(s_{i}, t_{q}) - T(s_{f}, t_{q})]\} \nonumber \\
&\times& \frac{(q_{i} + p_{i})_{\mu}}{(q_{i} + p_{i})\cdot K}
                                                        \label{eq:77}
\end{eqnarray}
\noindent in a manner analogous to Eqs.\ (\ref{eq:57}) and
(\ref{eq:58}). Here, we have used
the fact that $(q_{i} + p_{i})_{\mu}\epsilon^{\mu} = (q_{f} +
p_{f})_{\mu}
\epsilon^{\mu}$ and $(q_{i} + p_{i})\cdot K = (q_{f} + p_{f})\cdot
K$. [Again,
the amplitude $M^{I(TsTt)}_{\mu}$ vanishes in the C.~M. system and
the Coulomb
gauge since $M^{I(TsTt)}_{\mu}\epsilon^{\mu}$ is proportional to a
factor
$(q_{i} + p_{i})_{\mu}\epsilon^{\mu}$.] If we use Eq.\ (\ref{eq:52})
to expand all
T-matrices in Eqs.\ (\ref{eq:76}) and (\ref{eq:77}) about
($\bar{s}, \bar{t}$), then we can prove that
\begin{mathletters}
\begin{eqnarray}
M^{E(TsTt)}_{\mu} &=& M^{E(st)}_{\mu} + O(K)
                                                   \label{eq:78a}\\
M^{I(TsTt)}_{\mu} &=& M^{I(st)}_{\mu} + O(K) .
                                                   \label{eq:78b}
\end{eqnarray}
\end{mathletters}
\noindent Here, $\bar{s}$ and $\bar{t}$ are defined by
Eq.\ (\ref{eq:51}), $M^{E(st)}_{\mu}$ is the external term given
by Eq.\ (\ref{eq:57}), and $M^{I(st)}_{\mu}$ is the internal term
given by Eq.\ (\ref{eq:58}). Eqs.\ (\ref{eq:78a}) and (\ref{eq:78b})
show clearly that the amplitude $M^{Low(st)}_{\mu}$, derived
by using Low's standard procedure, is a special case [the first order
0($K^{0}$)
approximation] of the amplitude $M^{TsTts}_{\mu}$. In other words,
$M^{E(TsTt)}_{\mu}$ reduces to $M^{E(st)}_{\mu}$ and
$M^{I(TsTt)}_{\mu}$
reduces to $M^{I(st)}_{\mu}$ when T-matrices, $T(s_{x}, t_{Y}$), in
the
expression for $M^{E(TsTt)}_{\mu}$ and $M^{I(TsTt)}_{\mu}$ are
expanded about
($\bar{s}, \bar{t}$) and the O(K) term is neglected. It should be
emphasized
that if T-matrices $T(s_{x}, t_{y}$) vary rapidly with $s_{x}$ and/or
$t_{y}$
in the vicinity of a resonance, then the expansion of $T(s_{x},
t_{y}$) about
($\bar{s}, \bar{t}$), which is the essential step in the derivation
of the
amplitude $M^{Low(st)}_{\mu}$, is obviously not valid. In that case,
the
amplitude $M^{TsTts}_{\mu}$ which is free of $\partial T/\partial s$
and/or
$\partial T/\partial t$ is the only proper choice. In fact, the
result of
recent studies reveals that the amplitude $M^{TsTts}_{\mu}$ (or more
precisely
the special two-energy-two-angle amplitude $M^{TETAS}_{\mu}$) can be
used to
describe almost all the available $p^{12}C\gamma$ data (near both the
1.7 MeV and 0.5 MeV resonances) and $\pi^{\pm}p\gamma$ data [near the
$\Delta$(1232)
resonance]. These studies also show that the amplitude
$M^{Low(st)}_{\mu}$ has
failed to adequately describe both data.\\ (D) In this subsection, we
derive a
second general bremsstrahlung amplitude, in the soft-photon
approximation, for
a process whose elastic scattering is dominated by the diagrams shown
in Fig.~1(c). Since photon emission from the u-channel exchange
particles $F_{j}$, are involved, we choose the set of independent
Lorentz invariants which includes
$u_{j}(j = 1, 2)$, $t_{y}(y = p, q)$, and $\Delta_{z}(z = a, b, c, d)$.
The external emission amplitude is identical to the amplitude
$M^{E}_{\mu}(u, t,
\Delta$) given by Eq.\ (\ref{eq:60}). Since Fig.~1(a) is now dominated by
Fig.~1(c)
and Fig.~2 is dominated by Fig.~4, the internal emission amplitude
can be
written as
\begin{eqnarray}
M^{I(F)}_{\mu}(u, t, \Delta) &=& X_{a}\,T(u_{1}, t_{p}, \Delta_{a}) +
T(u_{2}, t_{p}, \Delta_{b}) X_{b}                       \nonumber \\
&+& X_{c}\,T(u_{2}, t_{q}, \Delta_{c}) + T(u_{1}, t_{q}, \Delta_{d})
X_{d},
                                                        \label{eq:79}
\end{eqnarray}
\noindent where $X_{z}(z = a, b, c, d)$ are the coefficients
to be specified. They can be uniquely determined if we demand that
$M^{I(F)}_{\mu}$ reduces to $\bar{M}^{I(F)}_{\mu}$ [given by
Eq.\ (\ref{eq:45}), with $T(u_{j}, t_{Y}, \Delta_{z})$
reduces to $\bar{T}(u_{j}, t_{Y})$, when Fig.~2(e) reduces to
Fig.~4(e). We find
\begin{eqnarray}
X_{a} &=& -Q_{A} \frac{(p_{i} - q_{f})_{\mu}}{(p_{i} - q_{f})\cdot
K},
                                                        \nonumber \\
X_{b} &=& Q_{A} \frac{(q_{i} - p_{f})_{\mu}}{(q_{i} - p_{f})\cdot K},
                                                        \nonumber \\
X_{c} &=& Q_{B} \frac{(q_{i} - p_{f})_{\mu}}{(q_{i} - p_{f})\cdot K},
                                                        \label{eq:80}
\end{eqnarray}
\noindent and
\begin{equation}
X_{d} = Q_{B} \frac{(p_{i} - q_{f})_{\mu}}{(p_{i} - q_{f})\cdot K}.
                                                        \nonumber
\end{equation}
\noindent Now, combining $M^{E}_{\mu}(u, t, \Delta)$ given by
Eq.\ (\ref{eq:60}) with $M^{I(F)}_{\mu}(u, t, \Delta)$ given
by Eqs.\ (\ref{eq:79}) and (\ref{eq:80}), we obtain
\begin{eqnarray}
M^{TuTt}_{\mu} &=& Q_{A} [ \frac{q_{f\mu}}{q_{f}\cdot K} -
\frac{(p_{i} -
q_{f})_{\mu}}{(p_{i} - q_{f})\cdot K} ] T(u_{1}, t_{p}, \Delta_{a})
                                                        \nonumber \\
&-& Q_{A}\,T(u_{2}, t_{p}, \Delta_{b})[ \frac{q_{i\mu}}{q_{i}\cdot K}
-
\frac{(q_{i} - p_{f})_{\mu}}{(q_{i} - p_{f})\cdot K} ]
                                                        \nonumber \\
&+& Q_{B} [ \frac{p_{f\mu}}{p_{f}\cdot K} - \frac{(q_{i} -
p_{f})_{\mu}}{(q_{i} - p_{f})\cdot K} ] T(u_{2}, t_{q}, \Delta_{c})
                                                       \label{eq:81}
\\
&-& Q_{B}\,T(u_{1}, t_{q}, \Delta_{d})[ \frac{p_{i\mu}}{p_{i}\cdot K}
-
\frac{(p_{i} - q_{f})_{\mu}}{(p_{i} - q_{f})\cdot K} ].
                                                          \nonumber
\end{eqnarray}
\noindent Again, no additional gauge term is required since
$M^{TuTt}_{\mu}$ is already gauge invariant; that is,
\[
M^{TuTt}_{\mu}D^{\mu} = 0.
\]
\noindent The amplitude $M^{TuTt}_{\mu}$ is an off-shell two-u-two-t
(TuTt)
amplitude which can be derived by using the modified Low procedure
for the A-B
bremsstrahlung process when internal emission from the u-channel
exchange
particles is important. To find an on-shell TuTt special amplitude
$M^{TuTts}_{\mu}$, we first expand $T(u_{j}, t_{Y}, \Delta_{z})$:
\begin{eqnarray}
T(u_{1}, t_{p}, \Delta_{a}) &=& T(u_{1}, t_{p}) + \frac{\partial
T(u_{1},
t_{p}, \Delta_{a})}{\partial \Delta_{a}} 2q_{f}\cdot K,
                                                        \nonumber \\
T(u_{2}, t_{p}, \Delta_{b}) &=& T(u_{2}, t_{p}) - \frac{\partial
T(u_{2},
t_{p}, \Delta_{b})}{\partial \Delta_{b}} 2q_{i}\cdot K,
                                                        \nonumber \\
T(u_{2}, t_{q}, \Delta_{a}) &=& T(u_{2}, t_{q}) + \frac{\partial
T(u_{2},
t_{q}, \Delta_{c})}{\partial \Delta_{c}} 2p_{f}\cdot K,
                                                     \label{eq:82}
\end{eqnarray}
\noindent and
\[
T(u_{1}, t_{q}, \Delta_{d}) = T(u_{1}, t_{q}) - \frac{\partial
T(u_{1},
t_{q}, \Delta_{d})}{\partial \Delta_{d}} 2p_{i}\cdot K,
\]
\noindent where
\begin{eqnarray}
T(u_{1}, t_{p}) &\equiv& T(u_{1}, t_{p}, m^{2}_{A}), \nonumber \\
T(u_{2}, t_{p}) &\equiv& T(u_{2}, t_{p}, m^{2}_{A}), \nonumber \\
T(u_{2}, t_{q}) &\equiv& T(u_{2}, t_{q}, m^{2}_{B}), \nonumber
\end{eqnarray}
\noindent and
\[
T(u_{1}, t_{q}) \equiv T(u_{1}, t_{q}, m^{2}_{B}).  \nonumber
\]
\noindent We then substitute Eq.\ (\ref{eq:80}) into
Eq.\ (\ref{eq:81}) to obtain
\begin{equation}
M^{TuTt}_{\mu} = M^{TuTts)}_{\mu} + M^{off(ut)}_{\mu},
                                                    \label{eq:83}
\end{equation}
\noindent where
\begin{eqnarray}
M^{TuTts}_{\mu} &=& Q_{A} [ \frac{q_{f\mu}}{q_{f}\cdot K} -
\frac{(p_{i} -
q_{f})_{\mu}}{(p_{i} - q_{f})\cdot K} ] T(u_{1}, t_{p})
                                                     \nonumber \\
&-& Q_{A}\,T(u_{2}, t_{p}) [ \frac{q_{i\mu}}{q_{i}\cdot K} -
\frac{(q_{i}
-p_{f})_{\mu}}{(q_{i} - p_{f})\cdot K} ]             \nonumber \\
&+& Q_{B} [ \frac{p_{f\mu}}{p_{f}\cdot K} - \frac{(q_{i} -
p_{f})_{\mu}}{(q_{i} - p_{f})\cdot K} ] T(u_{2}, t_{q})
                                                   \label{eq:84}\\
&-& Q_{B}\,T(u_{1}, t_{q}) [ \frac{p_{i\mu}}{p_{i}\cdot K} -
\frac{(p_{i}
-q_{f})_{\mu}}{(p_{i} - q_{f})\cdot K} ]              \nonumber
\end{eqnarray}
\noindent and $M^{off(ut)}_{\mu}$ includes those terms which involve
off-shell
derivatives of the T-matrix. Again, the off-shell amplitude
$M^{off(ut)}_{\mu}$
is ignored in the soft-photon approximation. The amplitude
$M^{TuTts}_{\mu}$ is
the on-shell TuTt special amplitude which should be used when
internal emission
from the u-channel exchange particles, F$_{j}$, are important.
It is easy to demonstrate that $M^{TuTts}_{\mu}$ given by
Eq.\ (\ref{eq:84}) is much more general than the amplitude
$M^{Low(ut)}_{\mu}$ given by Eq.\ (\ref{eq:67}). Again, we
divide the amplitude $M^{TuTts}_{\mu}$ into two parts, an external
term $M^{E(TuTt)}_{\mu}$ and an internal term $M^{I(TuTt)}_{\mu}$:

\begin{eqnarray}
M^{E(TuTt)}_{\mu} &=& Q_{A} \frac{q_{f\mu}}{q_{f}\cdot K}T(u_{1},
t_{p}) -
Q_{A}\,T(u_{2}, t_{p}) \frac{q_{i\mu}}{q_{i}\cdot K}
                                                        \nonumber \\
&+& Q_{B} \frac{p_{f\mu}}{p_{f}\cdot K}T(u_{2}, t_{q}) -
Q_{B}\,T(u_{1},
t_{q}) \frac{p_{i\mu}}{p_{i}\cdot K}                    \label{eq:85}
\end{eqnarray}
\noindent and
\begin{eqnarray}
M^{I(TuTt)}_{\mu} &=& -\{Q_{A}[T(u_{1}, t_{p}) - T(u_{2}, t_{p})] +
Q_{B}[T(u_{2}, t_{q}) - T(u_{1}, t_{q})]\}              \nonumber \\
&\times& \frac{(q_{i} - p_{f})_{\mu}}{(q_{i} - p_{f})\cdot K} .
                                                        \label{eq:86}
\end{eqnarray}
\noindent Here, we have used the following relation:
\[
\frac{(p_{i} - q_{f})_{\mu}\epsilon^{\mu}}{(p_{i} - q_{f})\cdot K} =
\frac{(q_{i} - p_{f})_{\mu}\epsilon^{\mu}}{(q_{i} - p_{f})\cdot K}.
                                                          \nonumber
\]
\noindent If we use Eqs.\ (\ref{eq:52}) and (\ref{eq:62}) to expand
all T-matrices in Eqs.\ (\ref{eq:85}) and (\ref{eq:86}) about
($\bar{u}, \bar{t}$), then we obtain
\begin{mathletters}
\begin{equation}
M^{E(TuTt)}_{\mu} = M^{E(ut)}_{\mu} + O(K)
                                                     \label{eq:87a}
\end{equation}
\noindent and
\begin{equation}
M^{I(TuTt)}_{\mu} = M^{I(ut)}_{\mu} + O(K).
                                                     \label{eq:87b}
\end{equation}
\end{mathletters}
\noindent Here, $M^{E(ut)}_{\mu}$ is the external term given by
Eq.\ (\ref{eq:68}), and $M^{I(ut)}_{\mu}$ is the internal term given
by Eq.\ (\ref{eq:69}), and we have used
the relation, $(q_{i} - q_{f})\cdot K = -(p_{i} - p_{f})\cdot K$.
Eqs.\ (\ref{eq:87a}) and (\ref{eq:87b}) demonstrate that
$M^{E(TsTt)}_{\mu}$ and $M^{I(TuTt)}_{\mu}$ reduce
to $M^{E(ut)}_{\mu}$ and $M^{I(ut)}_{\mu}$, respectively, if the
$T(u_{j},
t_{Y})$ in Eqs.\ (\ref{eq:85}) and (\ref{eq:86}) are expanded
about $(\bar{u},  \bar{t})$ and if O(K) terms are neglected.

  To summarize briefly, we have derived four soft-photon amplitudes,
$M^{Low(st)}_{\mu}(\bar{s},\bar{t}), \,M^{Low(ut)}_{\mu}(\bar{u},
\bar{t}), \,M^{TsTts}_{\mu}(s_{i},s_{f}; t_{p},t_{q})$, and
$M^{TuTts}_{\mu}(u_{1},u_{2}; t_{p},t_{q})$.
$M^{Low(st)}_{\mu}(\bar{s},\bar{t})$ and
$M^{Low(ut)}_{\mu}(\bar{u},\bar{t})$ were derived using Low's
standard procedure while
$M^{TsTts}_{\mu}(s_{i}, s_{f}; t_{p}, t_{q})$ and
$M^{TuTts}_{\mu}(u_{1}, u_{2}; t_{p}, t_{q})$ were derived using a
modified Low procedure. The amplitudes $M^{Low(st)}_{\mu}$ and
$M^{TsTts}_{\mu}$ depend on a
set of Lorentz invariants which include s and t. The amplitudes
$M^{Low(ut)}_{\mu}$ and $M^{TuTts}_{\mu}$, on the other hand, are
parametrized
in terms of Lorentz invariants u and t. In deriving
$M^{TsTts}_{\mu}(s_{i},
s_{f}; t_{p}, t_{q})$, we have imposed a condition that it reduce to
the
amplitude $\bar{M}^{TsTts}_{\mu}(s_{i}, s_{f}; t_{p}, t_{q})$, which
represents
photon emissions from a sum of one-particle t-channel exchange
diagrams and
one-particle s-channel exchange diagrams (the tree approximation).
Similarly,
in our derivation of the amplitude $M^{TuTts}_{\mu}(u_{1}, u_{2};
t_{p},
t_{q})$, we have imposed another condition that it reduce to the
amplitude
$\bar{M}^{TuTts}_{\mu}(u_{1}, u_{2}; t_{p}, t_{q})$, which represents
photon
emission from a sum of one-particle t-channel exchange diagrams and
one-particle u-channel exchange diagrams. Note that the expressions
for
$\bar{M}^{TsTts}_{\mu}$ and $\bar{M}^{TuTts}_{\mu}$ were derived in
last section by using the radiation decomposition identities
of Brodsky  and Brown.  We have proved that
$M^{Low(st)}_{\mu}$ and $M^{Low(ut)}_{\mu}$ can be
reproduced from $M^{TsTts}_{\mu}$ and $M^{TuTts}_{\mu}$,
respectively;
furthermore, the amplitudes $M^{TsTts}_{\mu}$ and $M^{TuTts}_{\mu}$
are the most general soft-photon amplitudes for hadron-hadron
bremsstrahlung processes which can be constructed by using the
modified Low procedure.  Finally, it is easy to show that the
amplitudes
$M^{Low(st)}_{\mu}$ and $M^{Low(ut)}_{\mu}$ and the amplitudes
$M^{TsTts}_{\mu}$ and $M^{TuTts}_{\mu}$ can be interchanged when
$p^{\mu}_{i},
p^{\mu}_{f}$ and $Q_{B}$ are replaced by $-p^{\mu}_{f}, -
p^{\mu}_{i}$ and
$-Q_{B}$, respectively. The relationships among the amplitudes
$\bar{M}^{TsTts}_{\mu}, \bar{M}^{TuTts}_{\mu}, M^{Low(st)}_{\mu},
M^{Low(ut)}_{\mu}, M^{TsTts}_{\mu}$ and $M^{TuTts}_{\mu}$ are
illustrated in
Fig.~5.

\section{DISCUSSION}\label{sec:5}
Six soft-photon amplitudes, $\bar{M}^{TsTts}_{\mu}$ [Eq.\ (\ref{eq:39})],
$\bar{M}^{TuTts}_{\mu}$ [Eq.\ (\ref{eq:46})], $M^{Low(st)}_{\mu}$
[Eq.\ (\ref{eq:56})], $M^{Low(ut)}_{\mu}$ [Eq.\ (\ref{eq:67})],
$M^{TsTts}_{\mu}$ [Eq.\ (\ref{eq:75})] and $M^{TuTts}_{\mu}$
[Eq.\ (\ref{eq:84})], have been derived in sections III and
IV. A primary purpose of this investigation is to explicate their
relationships and to explore their ranges of validity.
These six amplitudes can be divided into
two classes: (i) $\bar{M}^{TsTts}_{\mu}$, $M^{Low(st)}_{\mu}$ and
$M^{TsTts}_{\mu}$ as the first class [$M^{(1)}_{\mu}$ (s, t)] and
(ii)
$\bar{M}^{TuTts}_{\mu}$, $M^{Low(ut)}_{\mu}$ and $M^{TuTts}_{\mu}$ as
the second class [$M^{(2)}_{\mu}$ (u, t)].
As shown in Fig.~5, the following relationships have been
established: (A) $M^{TsTts}_{\mu}$ and
$M^{TuTts}_{\mu}$ reduce to $\bar{M}^{TsTts}_{\mu}$ and
$\bar{M}^{TuTts}_{\mu}$, respectively, in the tree level
approximation. (B) If
$\bar{M}^{TsTts}_{\mu}$ is expanded about ($\bar{s}, \bar{t}$) and
$\bar{M}^{TuTts}_{\mu}$ is expanded about ($\bar{u}, \bar{t}$),
assuming that
such expansions are valid, then the first two terms of the expansions
for
$\bar{M}^{TsTts}_{\mu}$ and $\bar{M}^{TuTts}_{\mu}$ give
$M^{Low(st)}_{\mu}$
and $M^{Low(ut)}_{\mu}$, respectively. (C) If
$p^{\mu}_{i}\longrightarrow
-p^{\mu}_{f}$, $p^{\mu}_{f}\longrightarrow - p^{\mu}_{i}$ and
$Q_{B}\longrightarrow -Q_{B}$, then $\bar{M}^{TsTts}_{\mu}
\longrightarrow
\bar{M}^{TuTts}_{\mu}$, $M^{Low(st)}_{\mu} \longrightarrow
M^{Low(ut)}_{\mu}$
and $M^{TsTts}_{\mu} \longrightarrow M^{TuTts}_{\mu}$, and vice
versa.
Now, let us consider the question about their ranges of validity.
Which
amplitude, $M^{TsTts}_{\mu}$ or $M^{TuTts}_{\mu}$, should be used to
describe a particular bremsstrahlung measurement?  The answer will
depend upon the nature of the bremsstrahlung process.  Let us examine
three cases:

(A) For a process whose elastic scattering is dominated by the tree
diagrams
shown in Fig.~1(b) or whose internal emission is dominated by the
diagrams
shown in Fig.~3(e), we must use the amplitude $M^{TsTts}_{\mu}$ for
bremsstrahlung calculations. That is, when the process is resonance
dominated, $M^{TsTts}_{\mu}$ is the correct choice.
Some well-known examples are the $\pi^{\pm} p\gamma$
process near the $\Delta$(1232) resonance, [9] the
p$^{12}C\gamma$ process near either the 1.7 MeV resonance or the 461
keV resonance, [7] and the p$^{16}O\gamma$ process near the 2.66 MeV
resonance [8].
These radiative processes have been systematically studied both
experimentally and theoretically. The following findings illustrate
why the
amplitude $M^{TsTts}_{\mu}$, not $M^{Low(st)}_{\mu}$, should be used
to
describe bremsstrahlung processes involving a resonance: (i) Using a
one-energy-two-angle amplitude, which is slightly different from the
amplitude
$M^{Low(st)}_{\mu}$, a UCLA group has calculated the $\pi^{\pm}
p\gamma$ cross
sections in order to compare with the cross sections measured by the
group [25]. The UCLA calculations have been repeated but using the
amplitude
$M^{Low(st)}_{\mu}$[14,15]. These two independent calculations
yield
essentially the same result. Typically, the calculated spectra at 298
MeV rise
steeply with increasing photon energy above K = 80 MeV in complete
disagreement
with the experimental data. The amplitude $M^{Low(st)}_{\mu}$ has
also been
used to calculate the $p^{12} C\gamma$ cross sections at 1.88 MeV for
a
scattering angle of 155$^{\circ}$[14,15].  The calculated cross
sections
show a large resonance peak around K = 270 keV in stark contrast with
the small
peak observed experimentally around K = 135 keV. In short, neither
$\pi^{\pm}
p\gamma$ nor the $p^{12} C\gamma$ data can be described by the
amplitude
$M^{Low(st)}_{\mu}$ or any other one-energy amplitude. These studies
also show
that the terms which involve $\partial T/\partial\bar{s}$ and
$\partial
T/\partial\bar{t}$ cause the problem. This is because the elastic
T-matrix,
which has been used as an input for bremsstrahlung calculations in
the
soft-photon approximation, varies rapidly with s and/or t in the
vicinity of a
resonance. In other words, the problem is directly related to the
invalid
expansions of the four half-off-shell T-matrices about ($\bar{s},
\bar{t}$) [or about ($s_{\alpha\beta}$, $t_{\alpha ' \beta '}$), where
$s_{\alpha\beta} = (\alpha s_{i} + \beta s_{f})/(\alpha + \beta)$
and $t_{\alpha ' \beta '} = (\alpha ' t_{p} + \beta '
t_{q})/(\alpha ' + \beta ')$], which are used in Low's standard
procedure for the derivation of $M^{Low(st)}_{\mu}$ and other
one-energy amplitudes. These expansions give rise to those
terms which depend upon
$\partial T/\partial\bar{s}$ and $\partial T/\partial\bar{t}$ in all
one-energy amplitudes.
(ii) From the amplitude $M^{TsTts}_{\mu}$ one may define an
amplitude designated the special two-energy-two-anble (TETAS)
amplitude $M^{TETAS}_{\mu}$, which is free of
$\partial T/\partial s$ and/or $\partial T/\partial t$.
The amplitude $M^{TETAS}_{\mu}$
has been thoroughly tested and has been found to describe the
data well for bremsstrahlung processes near a scattering resonance.
For example, $M^{TETASD}_{\mu}$ has been successfully applied to
extract the magnetic moments of the $\Delta^{++}$(1232) [9] and
$\Delta^{0}$(1232) [13] from the experimental $\pi^{\pm} p\gamma$
data and
$\pi^{-} p\gamma$ data, respectively. It is now well established that
this
amplitude can be used to describe almost all available $\pi^{12}
C\gamma$ and
$\pi^{\pm} p\gamma$ data. Furthermore, a direct, sensitive
experimental test of
various soft-photon amplitudes was made recently by the Brooklyn
group [7].  This test showed that the amplitude $M^{TETAS}_{\mu}$
provides an excellent
description of the $\pi^{12} C\gamma$ data not only in the
soft-photon region
but also in the hard-photon region.

(B) For a process whose elastic scattering is dominated by
the tree
diagrams shown in Fig.\ 1(c) or whose internal emission is dominated
by the
diagrams shown in Fig.\ 4(e), $M^{TuTts}_{\mu}$ should be used for
bremsstrahlung calculations. That is, when the process is exchange
current
dominated, $M^{TuTts}_{\mu}$ is optimal. An example of this is
neutron-proton
bremsstrahlung ($np\gamma$): (i) In the one-boson-exchange model, the
np
interaction involves the u-channel exchange of charged bosons.
(ii)  The $np\gamma$ cross section is dominated by the internal
emission from exchanged bosons.  More precisely, Brown and
Franklin have calculated the $np\gamma$ cross
sections using nonrelativistic potential model [11].  The
electromagnetic Hamiltonian used by these authors includes the
coupling of the
electromagnetic field to the nucleon currents $V^{1}_{em}$ and the
coupling of
the electromagnetic field to the exchange currents $V^{2}_{em}$. As a
result,
large exchange effects from $V^{2}_{em}$ were predicted. The
inclusion of the
$V^{2}_{em}$ term has been found to increase the $np\gamma$ cross
section by
about a factor of two. This finding has been confirmed very recently
by
Nakayama [12].   (iii) The $np\gamma$ cross sections at 200 MeV have
been calculated by Baier, Kuhnelt and Urban 26] using a
one-boson-exchange model and by Nyman [27] using a soft-photon
amplitude derived using Low's standard
procedure. The amplitude used by Baier et al.\ is equivalent to the
amplitude $\bar{M}^{TuTts}_{\mu}$ while the amplitude used by Nyman
is equivalent to $M^{Low(st)}_{\mu}$. When those two calculations
are compared, one can see that
the $np\gamma$ cross sections obtained by Baier et al.\ are
consistently a factor of 1.8 $\sim$ 2 times larger than those
obtained by Nyman. The obvious explanation of this result is that
the amplitude $M^{Low(st)}_{\mu}$ does not
contain any exchange effect since we have shown above that its
internal contribution is identically zero, while the amplitude
$\bar{M}^{TuTts}_{\mu}$
used by Baier et al.\ does include a nonzero internal contribution
from all exchanged bosons. [Note that the internal contribution
of the amplitudes
$M^{TsTts}_{\mu}$ and $\bar{M}^{Low(st)}_{\mu}$ involves a
factor of the form
$(q_{i} + p_{i})_{\mu} \epsilon^{\mu}$ which vanishes in the C.M.
system and in the Coulomb gauge.] Thus, the finding of Brown
and Franklin that the internal exchange contribution dominates
the $np\gamma$ cross section could also have been observed
by comparing the relativistic calculations of Nyman and
Baier et al..   The one-boson-exchange calculations of Baier
et al.\ are in much better agreement with the experimental data
of Brady and Young [28] than many other
calculations. This illustrates why the amplitude
$\bar{M}^{TuTts}_{\mu}$ (or $\bar{M}^{Low(ut)}_{\mu}$),
not the amplitude $\bar{M}^{TsTts}_{\mu}$  (or
$M^{Low(st)}_{\mu}$), should be used for $np\gamma$ calculations.

(C) For a process which involves little resonance effect
({\it i.~e.~,} it contains no resonant state or is observed in an
energy region far from resonance) and has very little contribution
from exchange effects (those due to
the u-channel exchange particles), we expect all six amplitudes
$\bar{M}^{TsTts}_{\mu}$, $M^{TsTts}_{\mu}$, $M^{Low(st)}_{\mu}$,
$\bar{M}^{TuTts}_{\mu}$, $M^{TuTts}_{\mu}$ and $M^{Low(ut)}_{\mu}$ to
yield
similar results, at least in the soft-photon region. This does not
mean that
they will give identical results but that the differences should not
be large.
A typical example is proton-proton bremsstrahlung ($pp\gamma$): (i)
As we have
already mentioned, there is no internal contribution from the
amplitudes
$\bar{M}^{TsTts}_{\mu}$, $M^{TsTts}_{\mu}$ and $M^{Low(st)}_{\mu}$
since it
vanishes in the C.M. system and in the Coulomb gauge. If
$M^{TsTts}_{\mu}$ is
expanded about ($\bar{s}, \bar{t}$), we obtain
\[
M^{TsTts}_{\mu} = M^{Low(st)}_{\mu} + O(K)
\]
which is exactly the sum of \ (\ref{eq:78a}) and (\ref{eq:78b}).
Here, O(K) involves the
derivatives of T-matrix with respect to $\bar{s}$ and $\bar{t}$. If
there is no
resonance effect, then derivatives of T with respect to $\bar{s}$ and
$\bar{t}$
will not produce significant structure and such an expansion is
valid. Hence,
the contribution from the O(K) term will be small, and we expect the
amplitudes
$M^{TsTts}_{\mu}$ and $M^{Low(st)}_{\mu}$ to give similar results.
(ii) For a
process which has very little contribution from exchange effects, the
amplitude
$\bar{M}^{TuTts}_{\mu}$ may be expanded about ($\bar{u}, \bar{t}$).
We find
\[
M^{TuTts}_{\mu} = M^{Low(ut)}_{\mu} + O(K)
\]
which is identical to the sum of Eqs.\ (\ref{eq:87a}) and (\ref{eq:87b}).
Again, if the derivatives of T with respect to $\bar{u}$ and $\bar{t}$
are small, then we expect that the contribution from the O(K) term
will be small. Therefore, the amplitudes $M^{TuTts}_{\mu}$ and
$M^{Low(ut)}_{\mu}$ should predict similar
cross sections. (iii) From Eq.\ (\ref{eq:69}), we can see that the internal
amplitude $M^{I(ut)}_{\mu}$ (of the amplitude $M^{Low(ut)}_{\mu}$)
contributes nothing if
$Q_{A} = Q_{B}$. Thus, like the amplitude $M^{Low(st)}_{\mu}$, there
is no internal contribution from $M^{Low(ut)}_{\mu}$ for the
$pp\gamma$ process. We therefore do not expect that the
$pp\gamma$ cross sections calculated using the
external part of the amplitude $M^{Low(st)}_{\mu}$ to be very
different from
those calculated using the external part of the amplitude
$M^{Low(ut)}_{\mu}$.
(iv) The $pp\gamma$ process has been extensively studied, both
experimentally
and theoretically, during the last three decades. Many different
calculations
(based on various models and approximations), including a soft-photon
approach
which uses an amplitude equivalent to $M^{Low(st)}_{\mu}$ and a
one-boson-exchange approach which uses an amplitude equivalent to
$\bar{M}^{TuTts}_{\mu}$, have been performed. The results of these
calculations
do differ, but their differences are indeed not large [29].  (v)
Since
two-nucleon interactions have been successfully described by the
one-boson-exchange model, we expect the difference between
$\bar{M}^{TuTts}_{\mu}$ and $M^{TuTts}_{\mu}$ to be small when these
amplitudes are applied to predict the $pp\gamma$ cross sections.

\section{SUMMARY AND CONCLUSIONS}\label{sec:6}

In conclusion, the primary purpose of this work is to point out
that there exist at least two independent classes of soft-photon
amplitudes, both of which are equally important for describing
hadron-hadron bremsstrahlung
processes. The two-s-two-t special amplitude
$M^{TsTts}_{\mu}$($s_{i}, s_{f};
t_{p}, t_{q}$), Eq.\ (\ref{eq:75}), is the general amplitude for
the first class, and this amplitude should be used to describe those
processes which are resonance
dominated. The two-u-two-t special amplitude
$M^{TuTts}_{\mu}$($u_{1}, u_{2};
t_{p}, t_{q}$), Eq.\ (\ref{eq:84}), is the general amplitude for
the second class, and it should be used to describe those processes
which are exchange current dominated. These two amplitudes can be
derived using a modified Low procedure, but not the standard (Low's
original) procedure.  The modified procedure involves one
additional step which allows us to take into account photon
emission from the internal line by imposing the condition that
$M^{TsTts}_{\mu}$ and $M^{TuTts}_{\mu}$ reduce to
$\bar{M}^{TsTts}_{\mu}$ and
$\bar{M}^{TuTts}_{\mu}$, respectively, at the tree level
approximation. The $\bar{M}^{TsTts}_{\mu}$
and $\bar{M}^{TuTts}_{\mu}$ amplitudes can be
rigorously derived from the relevant set of fundamental
bremsstrahlung diagrams
at the tree level, if we apply the radiation decomposition identities
of Brodsky and Brown to decompose the internal amplitude into
four quasi external amplitudes.

If $M^{TsTts}_{\mu}$ is expanded about ($\bar{s}, \bar{t}$) and
$M^{TuTts}_{\mu}$ is expanded about ($\bar{u}, \bar{t}$), assuming
that such
expansions are valid, the first two terms of the expansions yield
$M^{Low(st)}_{\mu}$($\bar{s}, \bar{t}$) and $M^{Low(ut)}_{\mu}$
($\bar{u},
\bar{t}$), respectively. Here, $M^{Low(st)}_{\mu}$ is a one-s-one-t
(or one-energy-one-angle) amplitude, a typical low amplitude
which can be derived
using the standard procedure. This amplitude has been regarded as the
sole soft-photon amplitude in the past, and it had been applied to
describe all possible bremsstrahlung processes without justification.
In addition to exploring
why $M^{Low(st)}_{\mu}$ cannot be used to describe processes
containing significant resonance effects, we also demonstrated why it
should fail to describe those processes with large exchange effects.
The amplitude $M^{Low(ut)}_{\mu}$, on the other hand,
is a one-u-one-t amplitude.    It is a new Low amplitude which
can also be derived by using the standard procedure. This
new amplitude has never before been studied.

We have demonstrated that we can transform the soft-photon amplitudes
in the
first class ($M^{TsTts}_{\mu}$, $\bar{M}^{TsTts}_{\mu}$,
$\bar{M}^{Low(st)}_{\mu}$) into the soft-photon amplitudes in the
second class
($M^{TuTts}_{\mu}$, $\bar{M}^{TuTts}_{\mu}$,
$\bar{M}^{Low(ut)}_{\mu}$) by
making the following variable transformations:
$p^{\mu}_{i}\leftrightarrow -
p^{\mu}_{f}$ and $Q_{B}\longrightarrow -Q_{B}$. This establishes the
relationship between the two independent classes.

Many amplitudes, especially those in the second class, discussed in
this work
are new.  Their ranges of validity and other properties are not well
understood.
Further systematic studies are required to understand these
amplitudes
thoroughly. These studies should include comparison with new
experimental work,
since the ultimate test of the utility of these soft-photon
amplitudes lies in
a comparison between the theoretical predictions and the experimenta
l data.
\pagebreak
\section{ACKNOWLEDGEMENTS}

The work of M.\ K.\ L.\ was supported in part by the City University
of New
York Professional Staff Congress-Board of Higher Education Faculty
Research
Award Program, while the work of B.\ F.\ G.\ was performed under the
auspices
of the U.\ S.\ Department of Energy.

\pagebreak

\newpage

\figure{1(a) Graphic representation of the A-B elastic scattering
process.  1(b) Feynman diagrams for the A-B elastic process at the
tree level.  The amplitude is approximated by a sum of
one-particle t-channel exchange  diagrams
(exchange of C$_{n}$ particles n = 1, 2, $\ldots$) and
one-particle s-channel exchange diagrams (exchange of D$_{\ell}$
particles, $\ell = 1, 2, \ldots$). 1(c) Feynman diagrams for the
A-B elastic process at the
tree level. The amplitude is approximated by a sum of one-particle
t-channel exchange diagrams (exchange of F$_{j}$ particles,
j = 1, 2, $\ldots$).\label{Fig.1}}

\figure{Feynman diagrams for bremsstrahlung: 2(a) - 2(d) are the
external emission diagrams; 2(e) is the internal emission diagrams.
These  diagrams are generated from the source graph,
Fig.\ 1(a).\label{Fig.2}}

\figure{Feynman diagrams for bremsstrahlung at the tree level:
3(a) - 3(d) are the external emission diagrams; 3(e) is the
internal emission  diagram. These diagrams are generated from the
source graphs, Fig.\ 1(b).\label{Fig.3}}

\figure{Same as Fig.\ 3, but the diagrams are generated from the
source graphs, Fig.\ 1(c).\label{Fig.4}}

\figure{Schematic representation of the relations among the six
soft-photon amplitudes derived in this work.  Five important
relations are shown here: (i) These six amplitudes can be divided
into two independent classes, ($M^{TsTts}_{\mu},
\bar{M}^{TsTts}_{\mu}, M^{Low(st)}_{\mu}$) as the  first
class and ($M^{TuTts}_{\mu}, \bar{M}^{TuTts}_{\mu},
M^{Low(ut)}_{\mu}$) as the second class. (ii) The general
amplitudes for the first and second classes are
$M^{TsTts}_{\mu}$ and $M^{TuTts}_{\mu}$, respectively. (iii) In the
tree level approximation, $M^{TsTts}_{\mu}$ reduces to
$\bar{M}^{TsTts}_{\mu}$ while $M^{TuTts}_{\mu}$ reduces to
$\bar{M}^{TuTts}_{\mu}$. (iv) When all T-matrices in
$M^{TsTts}_{\mu}$ are expanded about ($\bar{s}, \bar{t}$) and all
T-matrices in $M^{TuTts}_{\mu}$ are expanded about ($\bar{u},
\bar{t}$), then $M^{Low(st)}_{\mu}$ and $M^{Low(st)}_{\mu}$ can
be obtained. (v) The two classes of amplitude can be interchanged
($\bar{M}^{TsTts}_{\mu} \leftrightarrow \bar{M}^{TuTts}_{\mu},
M^{TsTts}_{\mu} \leftrightarrow M^{TuTts}_{\mu},
M^{Low(st)}_{\mu}\leftrightarrow M^{Low(ut)}_{\mu}$)
when $Q_B$ os replaced by $-Q_{B}$ ($Q_{B}\rightarrow -Q_{B}$)
and $p^{\mu}_{i}$ is interchanged with $-p^{\mu}_{f}$
$(p^{\mu}_{i}\leftrightarrow -p^{\mu}_{f})$.\label{Fig.5}}


\begin{references}
\bibitem{foot1} In general, all $\bar{T}$ matrices in Eq.\ (\ref{eq:26})
should be half-off-shell T-matrices. That is, we should have
$\bar{T}(s_{i}, t_{p},
\Delta_{a}), \bar{T}(s_{f}, t_{p}, \Delta_{b})$, $\bar{T}(s_{i},
t_{q},
\Delta_{c})$ and $\bar{T}(s_{f}, t_{q}, \Delta_{d})$ rather than
$\bar{T}(s_{i}, t_{p}), \bar{T}(s_{f}, t_{p}), \bar{T}(s_{i}, t_{q})$
and
$\bar{T}(s_{f}, t_{q})$, respectively, to indicate that one of the
four legs is
off its mass shell. We have ignored the $\Delta_{z}(z = a, b, c, d)$
dependence
here simply because none of the $\bar{T}$ matrices explicitly depend
upon
$\Delta_{z}$ in this special case.
\bibitem{foot2} As indicated in \cite{foot1}, in general all
$\bar{T}$ matrices
in Eq.\ (\ref{eq:39}) may depend not only on $s_{x}(x = i, j)$
and $t_{y}(y = p, q)$ but
also on $\Delta_{z}(z = a, b, c, d)$. In that case, the amplitude
given by Eq.\ (\ref{eq:39}) should be classified as the two-s-two-t
amplitude  $M^{TsTt}_{\mu}$. If
$M^{TsTt}_{\mu}$ is expanded and all off-shell terms are ignored,
then the
first two terms of the expansion define the amplitude
$M^{TsTts}_{\mu}$. The
expression for $M^{TsTts}_{\mu}$ will be exactly the same as that
given by Eq.\ (\ref{eq:39}).
\bibitem{foot3} In this special case, none of the $\bar{T}$ matrices
in Eq.\ (\ref{eq:41})
depend on $\Delta_{z}(z = a, b, c, d)$ even though $\bar{T}$ may
depend on
$\Delta_{z}$ in general; see \cite{foot1}.
\bibitem{foot4} If all $\bar{T}$ matrices in Eq\ (\ref{eq:46}) depend on
$\Delta_{z}(z = a, b,
c, d)$, then the amplitude given by Eq.\ (\ref{eq:46}) should be
classified as the
two-u-two-t amplitude $M^{TuTt}_{\mu}$. Expanding $M^{TuTt}_{\mu}$
and
neglecting all off-shell terms, the amplitude $M^{TuTts}_{\mu}$,
which has
exactly the same expression given by Eq.\ (\ref{eq:46}) (i.\ e.\ , no
$\bar{T}$ matrix
depends on $\Delta_{z}$), is defined by the first two terms of the
expansion.
\end{references}
\end{document}